\documentclass[aps,prb,twocolumn,superscriptaddress]{revtex4}
\usepackage{amsmath}
\usepackage{amssymb}
\usepackage{epsfig}
\usepackage{graphicx}

\begin{document}

\newcommand{\bra}[1]{\langle {#1} |}
\newcommand{\ket}[1]{| {#1} \rangle}
\newcommand{\expect}[1]{\langle {#1} \rangle}
\newcommand{\ketn}[1]{ {#1} \rangle}
\newcommand{\bea}{\begin{eqnarray}}
\newcommand{\eea}{\end{eqnarray}}
\newcommand{\beq}{\begin{equation}}
\newcommand{\eeq}{\end{equation}}
\newcommand{\lav}{\langle}
\newcommand{\rav}{\rangle}
\def \tr{{\mbox{tr~}}}
\def \ra{{\rightarrow}}
\def \ua{{\uparrow}}
\def \da{{\downarrow}}
\def \be{\begin{equation}}
\def \ee{\end{equation}}
\def \ba{\begin{array}}
\def \ea{\end{array}}
\def \bea{\begin{eqnarray}}
\def \eea{\end{eqnarray}}
\def \nn{\nonumber}
\def \l{\left}
\def \r{\right}
\def \half{{1\over 2}}
\def \etal{{\it {et al}}}
\def \cH{{\cal{H}}}
\def \cM{{\cal{M}}}
\def \cN{{\cal{N}}}
\def \cQ{{\cal Q}}
\def \cI{{\cal I}}
\def \cV{{\cal V}}
\def \cG{{\cal G}}
\def \cF{{\cal F}}
\def \cZ{{\cal Z}}
\def \bS{{\bf S}}
\def \bI{{\bf I}}
\def \bL{{\bf L}}
\def \bG{{\bf G}}
\def \bQ{{\bf Q}}
\def \bK{{\bf K}}
\def \bR{{\bf R}}
\def \br{{\bf r}}
\def \bu{{\bf u}}
\def \bq{{\bf q}}
\def \bk{{\bf k}}
\def \bv{{\bf v}}
\def \bx{{\bf x}}
\def \bpsi{{\bar{\psi}}}
\def \tJ{{\tilde{J}}}
\def \W{{\Omega}}
\def \e{{\epsilon}}
\def \lam{{\lambda}}
\def \L{{\mathcal L}}
\def \a{{\alpha}}
\def \t{{\tau}}
\def \b{{\beta}}
\def \g{{\gamma}}
\def \D{{\Delta}}
\def \de{{\delta}}
\def \w{{\omega}}
\def \r{{\rho}}
\def \s{{\sigma}}
\def \f{{\varphi}}
\def \x{{\chi}}
\def \e{{\epsilon}}
\def \h{{\eta}}
\def \G{{\Gamma}}
\def \z{{\zeta}}
\def \hatt{{\hat{\t}}}
\def \hn{{\bar{n}}}
\def \vk{{\bf{k}}}
\def \vq{{\bf{q}}}
\def \gk{{\g_{\vk}}}
\def \d{{d\!\!\!^-}}
\def \ket#1{{\,|\,#1\,\rangle\,}}
\def \bra#1{{\,\langle\,#1\,|\,}}
\def \braket#1#2{{\,\langle\,#1\,|\,#2\,\rangle\,}}

\def \nd{^{\vphantom{\dagger}}}
\def \yd{^\dagger}
\def \av#1{{\langle#1\rangle}}
\def \journal#1#2#3#4{{#1\ {\bf #2}, #3 (#4)}} 
\def \fig#1{{Fig.~\ref{#1}}}
\def \eqn#1{{Eqn.~(\ref{#1})}}
\def \tbl#1{{Table~\ref{#1}}}
\def \app#1{{Appendix~\ref{#1}}}
\def \Ref#1{{Ref.~\onlinecite{#1}}}
\def \sec#1{{Section~\ref{#1}}}

\def \im{i} 
\def \idmat{1} 
\def \dif{d} 
\def \mcD{\mathcal{D}} %
\def \cellno{N} 
\def \veck{{\bf k}} 
\def \vecq{{\bf q}} 
\def \vecR{{\bf R}} 
\def \vecr{{\bf r}} 
\def \vece{{\bf e}} 
\def \vecS{{\bf S}} 
\def \chir{\eta} 
\def \flucy{\epsilon^{y}} 
\def \flucz{\epsilon^{z}} 
\def \rfy{\tilde{\epsilon}^{y}} 
\def \rfz{\tilde{\epsilon}^{z}} 
\def \tilH{\tilde{H}} 
\def \tilS{\tilde{S}} 
\def \tilb{\tilde{\beta}} 
\def \lgi{\lambda} 
\def \lgr{\mu} 
\def \tillg{\tilde{\lambda}} 

\title{Quantum and Classical Spins on the Spatially Distorted Kagom\'e Lattice:
 Applications to volborthite Cu$_3$V$_2$O$_7$(OH)$_2$ $\cdot$ 2 H$_2$O}
\author{Fa Wang}
\affiliation{Material Sciences Division, Lawrence Berkeley
Laboratories, Berkeley, California 94720}
\affiliation{Department of Physics, University of California, Berkeley,
California 94720}
\author{Ashvin Vishwanath}
\affiliation{Material Sciences Division, Lawrence Berkeley
Laboratories, Berkeley, California 94720}
\affiliation{Department of Physics, University of California, Berkeley,
California 94720}
\author{Yong Baek Kim}
\affiliation{Department of Physics, University of Toronto, Toronto, Ontario M5S 1A7, Canada.}
\affiliation{Kavli Institute for Theoretical Physics, University of California,
Santa Barbara, California 93106}

\date{\today}

\begin{abstract}
In Volborthite, spin-1/2 moments form a distorted Kagom\'e lattice,
of corner sharing isosceles triangles with exchange constants $J$ on
two bonds and $J'$ on the third bond. We study the properties of
such spin systems, and show that despite the distortion, the lattice
retains a great deal of frustration. Although sub-extensive, the
classical ground state degeneracy remains very large, growing
exponentially with the system perimeter. We consider degeneracy
lifting by thermal and quantum fluctuations. To linear (spin wave)
order, the degeneracy is found to stay intact. Two complementary
approaches are therefore introduced, appropriate to low and high
temperatures, which point to the same ordered pattern for $J'>J$. In
the low temperature limit, an effective chirality Hamiltonian is
derived from non-linear spin waves which predicts a transition on
increasing $J'/J$, from $\sqrt 3\times \sqrt 3$ type order to a new
ferrimagnetic {\em striped chirality} order with a doubled unit
cell. This is confirmed by a large-$n$ approximation on the O($n$)
model on this lattice. While the saddle point solution produces a
line degeneracy, $O(1/n)$ corrections select the non-trivial
wavevector of the striped chirality state. The quantum limit of spin-1/2 
on this lattice is studied via exact small system
diagonalization and compare well with experimental results at
intermediate temperatures. We suggest that the very low temperature
spin frozen state seen in NMR experiments may be related to the
disconnected nature of classical ground states on this lattice,
which leads to a prediction for NMR line shapes.
\end{abstract}


\maketitle
\date{\today}

\section{Introduction}
The study of frustrated magnetic insulators has witnessed a
resurgence in recent times, with the discovery of a number of
interesting materials with frustrated spin interactions. Amongst the
most geometrically frustrated lattices are the pyrochlore and the
Kagom\'e lattice, and perhaps the most interesting class of systems
are those that combine strong quantum fluctuations with frustration.
Recently, spin-1/2 systems on the Kagom\'e lattice have been
identified, the mineral Volborthite ${\rm Cu_3V_2O_7(OH)_2\cdot
2H_2O}$ \cite{Hiroi:} and Herbertsmithite \cite{Nocera}. In the
former the equilateral Kagom\'e triangles are distorted into
isoceles triangles, rendering two of the nearest neighbor exchange
constants different from the third. In the latter case, the Kagom\'e
lattice is believed to be structurally perfect, but with perhaps a
small percentage of impurity spins. Nevertheless, both systems
display low temperature physics very different from their
unfrustrated counterparts, and do not show signs of ordering down to
temperatures well below the exchange coupling strength.

While a lot of theoretical effort has gone into characterizing the
ideal frustrated lattices, distortions of the ideal structure,
although common, have received less attention\cite{tchernyshyov}. In
many frustrated magnets, lattice distortions occur spontaneously to
relieve the frustration, leading to a strong coupling between
magnetic and structural order parameters. Such `multi-ferroic'
couplings are highly prized from the technological viewpoint for
convenient manipulation of magnetism \cite{Ramesh_Spaldin} and
certain frustrated magnets are natural candidates\cite{Mostovoy}.
This provides further motivation for studying the effect of
distortions. From the theoretical viewpoint, the partial lifting of
degeneracy from distortions can lead to a more tractable level of
frustration, and allow for new theoretical approaches. Here, we
consider the effect of lattice distortion on the Kagom\'e lattice.
The class of lattice distortions we focus on are motivated by the
material Volborthite, whose structure consists of corner sharing
isoceles triangles. Bonds along two directions then have exchange
constant $J$ while the bond along the third direction has exchange
constant $J'=\alpha J$. {In Volborthite, it is not definitively
known if $\alpha>1$ or $\alpha<1$, although a comparison of bond
lengths seems to favor the former\cite{Lafontaine}. Hence we treat both kinds of
anisotropy in this paper, with slightly more emphasis on the
$\alpha>1$ case.}

We attack this problem first from the classical zero temperature
limit.
We show that for a wide range of distortions, the large classical
degeneracy of the Heisenberg model on the isotropic Kagom\'e lattice
is partially lifted, and the number of coplanar ground states now
scales in a sub-extensive fashion, as the exponential of the linear
system size. An interesting comparison here is with the isotropic
Kagom\'e and pyrochlore lattices, where the extensively degenerate
ground state can be specified in terms of local constraints
reminiscent of the Gauss law of a lattice gauge
theory\cite{youngblood_axe,harris}. In fact, that analogy has been
carried further to describe new quantum phases of frustrated magnets
corresponding to the coulomb phase of the lattice gauge theory
\cite{balents_fisher_girvin,hermele}. In contrast, the subextensive
classical degeneracy of the distorted Kagom\'e lattice is naturally
thought of as arising from constrains on one dimensional structures,
and the `soft-spin' dispersion on this lattice features a line
degeneracy in the Brillouin zone, unlike the flat band of the
Kagom\'e lattice. Both these features are shared by pure ring
exchange models on the square lattice as studied in
\Ref{paramekanti_balents_fisher}, where a new spin
liquid phase, the excitonic bose liquid, was discussed.

In contrast to the isotropic Kagom\'e system, the ground states of
the distorted Kagom\'e lattice are not connected by local moves,
requiring moving an infinite number of spins  to make transitions
from one configuration to another. We suggest that this difference
may be related to the experimental observation of spin freezing seen
in NMR experiments at low temperatures in Volborthite (but not in
the isotropic Kagom\'e compound Herbertsmithite). The classical
ground state ensemble may then be expected to capture aspects of
this glassy state, which we use to make experimental predictions.

Next, we consider the question - if a spin system on this lattice
develops long range magnetic order, what is the preferred spin
pattern? The degeneracy is expected to be broken by fluctuation
effects, and hence we analyze the effect of quantum and thermally
excited spin waves in the harmonic approximation. Remarkably, the
spin waves are found to have a precisely flat dispersion, as in the
ideal Kagom\'e case, and do not distinguish between the classical
ground states at this level. To proceed we consider thermal
fluctuations in the classical model with $\alpha > 1$ in two
complementary ways, approaching from high and low temperatures.
These are found to be consistent with one another and point to a new
ferrimagnetic state, characterized by alternating chirality stripes,
and a doubled unit cell, which we call the {\em chirality stripe
state}. The first calculation consists of combining the low
temperature non-linear spin wave expansion with the effective
chirality Hamiltonian technique pioneered by Henley \cite{Henley:}.
While at the isotropic point our method picks out the
$\sqrt{3}\times\sqrt{3}$ state, consistent with many other studies
\cite{Chubukov:,Chandra:,Henley:,Huse_Rutenberg:,Sachdev:}, turning
up the spatial anisotropy leads to a transition into a new state -
the chirality stripe state. To attack the problem from the opposite,
disordered limit, we consider generalization to the classical O($n$)
model which is tractable in the limit $n\rightarrow \infty$, and
captures the fluctuating nature of the spins at high temperatures.
At the saddle point level, the flat band degeneracy of the ideal
Kagom\'e case is shrunk down to a line degeneracy for $\alpha>1$.
Fully lifting the degeneracy requires going to the next order in
$1/n$, which we accomplish by utilizing the high temperature
expansion. The selected state has the same nontrivial wavevector as
the chirality stripe state providing additional confirmation. {
In contrast, when $\alpha<1$, the large $n$ saddle point itself
picks out the $q=0$ wavevector.}

Finally, we study the problem in the quantum limit, via exact
diagonalization studies on small (12-site) systems with spin-1/2.
Bulk properties such as specific heat and magnetic susceptibility
at intermediate to high temperatures are found to be rather insensitive
to the anisotropy and differences arise only below temperatures of
about $J/5$,
as seen in experiments \cite{Hiroi:}. On the other hand, the ground
state of the small cluster is found to be a spin singlet
and {the spin gap
decreases on increasing anisotropy}.

The structure of this paper is as follows. In \sec{sec:groundstates} 
we discuss the classical ground states of the
distorted Kagom\'e model as well as the properties of the ground
state ensemble, and possible connections to the NMR experiments on
the low temperature state in Volborthite. Next, we address the
question of which spin ordered pattern is favored by fluctuations on
this lattice using two approaches, first by deriving an effective
chirality Hamiltonian from non-linear spin waves in \sec{sec:orderbydisorder} 
and next via a classical large-n O($n$)
approach, in \sec{sec:largen}, which produce consistent
results. Finally, the problem is treated in the extreme quantum
limit via exact diagonalization of small systems in
\sec{sec:quantumstudy}. Details of calculations are relegated
to three appendices.

\section{Classical Ground States}\label{sec:groundstates}
\begin{figure}
\includegraphics{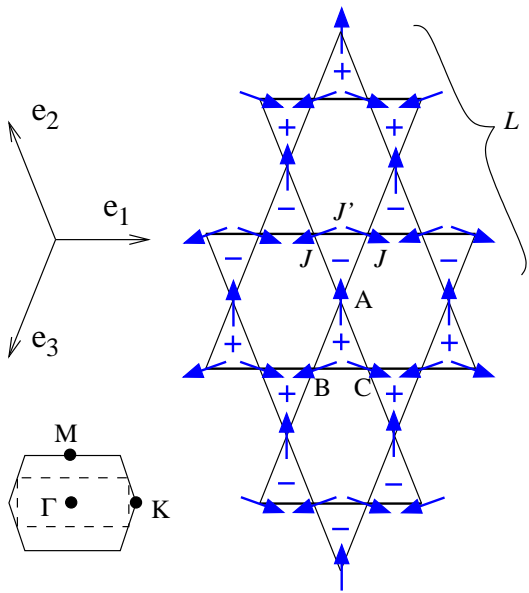}
\caption{(Color online) a $L\times L$ ($L=2$) distorted Kagom\'e lattice.
$\vece_1,\vece_2,\vece_3$ are three lattice translation vectors. The
exchange constant $J$ for bonds along the $\vece_2$ and $\vece_3$
directions are equal, but different from $J'=\alpha J$, the exchange
constant for bonds along the $\vece_1$ direction. For Volborthite,
it is believed that $\alpha>1$. $\Gamma,{\rm M},{\rm K}$ are high
symmetry points in the Brillouin zone (BZ). The proposed
spin-ordered state has alternating positive/negative chirality
stripes and Fourier components at $\Gamma$- and M-points in BZ.
Dashed rectangle is the reduced BZ for the doubled magnetic unit
cell.} \label{fig:lattice}
\end{figure}

Consider the antiferromagnetic Heisenberg model on the distorted
Kagom\'e lattice (\fig{fig:lattice}) with different couplings for
bonds on the three principal directions, \be
\begin{split}
H=&\sum_{\rm triangles}{ (J_{\rm AB}\vecS_{\rm A}\cdot\vecS_{\rm B}
+J_{\rm BC}\vecS_{\rm B}\cdot\vecS_{\rm C}+J_{\rm CA}\vecS_{\rm C}\cdot\vecS_{\rm A}) }\\
=&\frac{\prod{J}}{2}\sum_{\rm triangles}
{\left ( \frac{\vecS_{\rm A}}{J_{\rm BC}}+\frac{\vecS_{\rm B}}{J_{\rm CA}}
+\frac{\vecS_{\rm C}}{J_{\rm AB}} \right )^2}-{\rm constant}
\end{split}
\nn \ee where $\vecS$ are quantum or classical spins, $\prod{J}$
means $J_{\rm AB} J_{\rm BC} J_{\rm CA}$, and ${\rm A,B,C}$ are
indices for the three sublattices.

If $J_{\rm AB},\,J_{\rm BC},\,J_{\rm CA}$ are all different, we call
the lattice {\em fully distorted Kagom\'e lattice}. In this paper
however we consider mainly the {\em distorted Kagom\'e lattice} in
which $J_{\rm AB}=J_{\rm CA}\neq J_{\rm BC}$. For simplicity we set
$J_{\rm AB}=J_{\rm CA}=1$ and $J_{\rm BC}=\alpha$. The Hamiltonian
simplifies to the following form,
\be
\begin{split}
H&=\sum_{\rm triangles}{ (\vecS_{\rm A}\cdot\vecS_{\rm B}
+\alpha\vecS_{\rm B}\cdot\vecS_{\rm C}+\vecS_{\rm C}\cdot\vecS_{\rm A}) }\\
&=\frac{\alpha}{2}\sum_{\rm triangles}{\left [(1/\alpha)\vecS_{\rm A}+\vecS_{\rm B}
+\vecS_{\rm C}\right ]^2-{\rm constant}}
\end{split}
\label{equ:H}
\ee
There are two simple limits. In one limiting case,
$\alpha\rightarrow 0$, the lattice becomes a decorated square
lattice, with additional sites at the midpoints of square lattice
edges. In the other, quasi-1D, limit $\alpha\rightarrow \infty$ the
lattice turns into decoupled antiferromagnetic chains and `free'
spins.

From the lattice structure\cite{Lafontaine}, especially the Cu-O
bond lengths data, of Volborthite we expect that $\alpha>1$ in that
material, although there is no direct experimental data available or
quantitative first principles calculations available yet. Hence, the
$\alpha<1$ case is also considered in some of the following
theoretical treatments.

\begin{figure}
\includegraphics{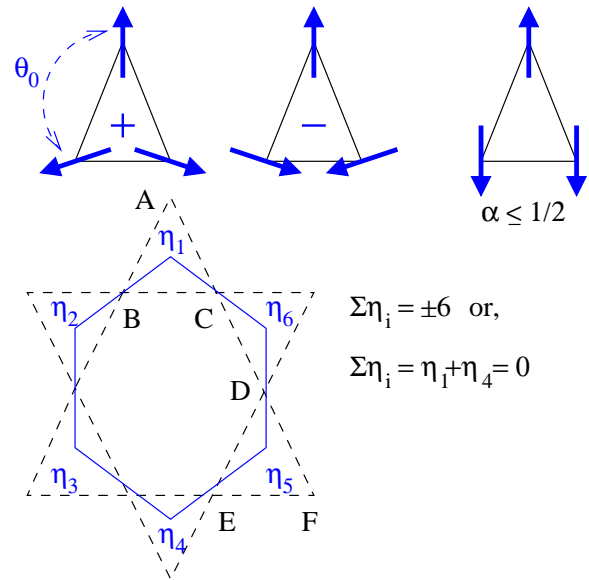}
\caption{(Color online) ground states of a single triangle
($\theta_0=\arccos(-1/2\alpha)$), definition of chirality variables,
and constraint on the six chirality variables for the distorted
Kagom\'e model, on a single hexagon of the honeycomb chirality
lattice. ABCDEF are labels of six spin sites used to calculate
the effective chirality interactions in a later section.
}
\label{fig:trigGSconstraints}
\end{figure}

The first step of studying the classical ground states on the
lattice is to solve the classical ground states of a single
triangle. Setting the `cluster spin' in \eqn{equ:H} to zero we can
solve the angle between A-site spin and B(C)-site spin, denoted as
$\theta_0=\arccos(-1/2\alpha)$ (see \fig{fig:trigGSconstraints}).
Since $\alpha\neq 1$, this angle will be in general incommensurate
to $2\pi$. We ignore the accidental commensurate cases in the
following discussion since they form a measure-zero set of $\alpha$.
Then the (3-state) Potts model description for the coplanar ground
states of the isotropic Kagom\'e case does not work for the
distorted Kagom\'e lattice.

A special case is $\alpha < 1/2$.
In this case there is no way that the `cluster spin' can be zero
and the classical ground state is a collinear state
with A-site spin anti-parallel to B(C)-site spin (\fig{fig:trigGSconstraints}).
Thus, for $\alpha \le 1/2$, the classical ground state is collinear and
there is no degeneracy except a global spin rotation.
Notice that the lattice becomes bipartite (not frustrated) in the
limit $\alpha = 0$.
This classical consideration shows that the frustration of BC bonds is
ineffective for nonzero $\alpha\leq 1/2$. Later we will see from exact
diagonalization study that this naive classical picture survives
in quantum regime. The classical collinear state has a macroscopic
net moment and is a `ferrimagnetic' state.

For $\alpha > 1/2$ case we expect that coplanar classical ground
states are favored by thermal or quantum fluctuations, and there
will be zero energy band(s) for the O($n$) model with $n\geq 3$,
because the Hamiltonian in \eqn{equ:H} can be written as a sum of
squares of `cluster spins'\cite{Moessner:}. Then it is convenient to
utilize the chirality variables used in the isotropic Kagom\'e
model\cite{Chandra:}. The chirality variables are Ising variables
living at the centers of triangles, thus forming a honeycomb
lattice. The positive or negative chirality variable represents the
cluster of three spins on a triangle rotating counter-clockwise or
clockwise when one goes from A- to B- then to C-site, or $\vecS_{\rm
A}\times(\vecS_{\rm B}-\vecS_{\rm C})$ pointing toward the $+z$ or
$-z$ direction, assuming all spins lie in the $x-y$ plane
(\fig{fig:trigGSconstraints}).

It should be emphasized that the chirality variables are {\em not}
independent. They determine how spin rotates (counter-clockwise or
clockwise) when one walks along a bond, but after walking along a
closed loop on the lattice the spin should go back to the initial
direction. We need only to consider length-six hexagonal loops on
the (distorted) Kagom\'e lattice. Each one of these loops will
impose a constraint on the six chirality variables $\chir$ in the
corresponding hexagon in the honeycomb chirality lattice
(\fig{fig:trigGSconstraints}), \be
\chir_1(2\theta_0)-\chir_2\theta_0-\chir_3\theta_0+\chir_4(2\theta_0)
-\chir_5\theta_0-\chir_6\theta_0=0\mod 2\pi \ee
For the isotropic Kagom\'e antiferromagnet, $\theta_0=2\pi/3$ and
the constraint simplifies to $\sum_{i=1}^{6}{\chir_i}=\pm 6\ {\rm
or}\ 0$. There are 22 allowed patterns on a single hexagon out of
$2^6=64$ combinations. For the distorted Kagom\'e model, $\theta_0$
is incommensurate to $2\pi$ and the constraint is more restrictive:
$\sum_{i=1}^{6}{\chir_i}=\pm 6$ or, $\sum_{i=1}^{6}{\chir_i}=0\ {\rm
and}\ \chir_1+\chir_4=0$. The last equation is the new constraint
compared to the isotropic Kagom\'e lattice. Note, this constraint
holds for all $\alpha \neq 1$, so long as a coplanar ground state is
favored, i.e. $\alpha > 1/2$. There are only 14 allowed patterns on a
single hexagon. For the fully distorted Kagom\'e lattice, the
constraint is even more restrictive: $\sum_{i=1}^{6}{\chir_i}=\pm 6$
or, $\chir_1+\chir_4=\chir_2+\chir_5=\chir_3+\chir_6=0$. There are
only 10 allowed patterns on a single hexagon.

\subsection{Properties of Coplanar Ground States on Distorted Kagom\'e Lattice}
The degree of degeneracy for these models on a lattice is a much
more subtle problem. From Baxter's solution\cite{Baxter:} we know
that the degeneracy of the coplanar ground states of the isotropic
Kagom\'e antiferromagnet (or 3-state Potts antiferromagnet) is
extensive, $\exp(0.379\cellno)$, where $\cellno$ is the number of
Kagom\'e unit cells.

By counting the allowed chirality patterns for the distorted
Kagom\'e model with the $L\times L$ open boundary geometry in
\fig{fig:lattice} up to $L=9$, we conclude that the degeneracy is
`sub-extensive', about $\exp(2.2 L)$. \tbl{table:enum} lists the
exact enumeration result.

\app{app:Tmatrix} derives the asymptotic formula of the degeneracy by
transfer matrix method for a slightly different geometry with
periodic boundary condition. The `sub-extensive' behavior is proved by
rigorous upper and lower bounds and the asymptotic formula.

For fully distorted Kagom\'e model the degeneracy is also
`sub-extensive', about $\exp(1.4 L)$ for the geometry in
\fig{fig:lattice}.

One should be aware that the constant in the exponent depends on geometry and
boundary conditions. Notice that \app{app:Tmatrix}\ uses another geometry
so that the result is not exactly the same as the enumeration results,
although they both show `sub-extensive' behavior.

Another issue about classical degeneracy is the existence of the
so-called `weather-vane' modes. In the isotropic Kagom\'e model
those local zero-energy modes were argued to favor the
$\sqrt{3}\times\sqrt{3}$ state\cite{Chandra:}. However one can
easily prove that in the distorted Kagom\'e O(3) model there is no
{\em local} `weather-vane' modes. This is because the cluster of
spins of a `weather-vane' mode must be bounded by spins pointing to
the same direction. Those boundary spins inevitably involves all
three sublattices if the cluster is finite. But an A-site spin can
never be in the same direction as a B-site spin if
$\theta_0=\arccos(-1/2\alpha)$ is incommensurate to $2\pi$.

There could still be non-local `weather-vane' modes involving an
infinite number of spins in the thermodynamic limit. But the number
of these modes do not scale as the area of the system. In this
respect, the ground state manifold of the distorted Kagom\'e model
is much less connected than that of the isotropic Kagom\'e model.
Thus glassy behavior is more likely to happen in the distorted
model.

Huse and Rutenberg studied the ground state ensemble of the
isotropic Kagom\'e antiferromagnet\cite{Huse_Rutenberg:} by field
theoretical and Monte Carlo methods, and found that the spin-spin
correlation has the $\sqrt{3}\times\sqrt{3}$ state signature but
with power-law decay.

We study the classical ground state ensemble of the distorted model
by measuring the ensemble averaged spin-spin correlation. Lacking a
good Monte Carlo algorithm we use the exact enumeration result for
$L\times L$ lattice with open boundary up to $L=9$. Because of the
small size and possible boundary effects we have not been able to
extract the scaling form of the correlation functions. However the
result is qualitatively different from those of the isotropic
Kagom\'e antiferromagnet. For A-sublattice the correlation has a
large $q=0$ ($\Gamma$-point) component. For B(C)-sublattice the
correlation has a large Fourier component at the M-point, the
mid-point of the BZ top(bottom) edge.

Based on these hints we propose an ordering pattern as in \fig{fig:lattice}.
It has horizontal alternating stripes of positive(negative) chiralities.
We will later call it the {\em chirality stripe state}.
This pattern doubles the magnetic unit cell in the vertical direction,
thus reduces the BZ, and the M-point is actually equivalent to
the $\Gamma$-point for the reduced BZ (\fig{fig:lattice}).

To further confirm this we measured the mean-square of three Fourier modes of
the chirality variables $\av{m^2}$:\\
(i) the uniform pattern, corresponding to the $q=0$ ($\Gamma$-point in BZ)
spin configuration, with $m_\Gamma=\sum{\chir}$; (ii) the staggered
pattern, corresponding to the $\sqrt{3}\times\sqrt{3}$ spin
configuration of the isotropic case or K-point in BZ, with
$m_{\rm K}=\sum{\pm \chir}$ where the two sublattices in the honeycomb
chirality lattice have opposite $\pm$ sign; and (iii) the chirality
stripe pattern, corresponding to our proposed spin configuration
(M-point in BZ), with $m_{\rm M}=\sum{\pm \chir \exp(\im \veck_{\rm
M}\cdot\vecR)}$ where the $\pm$ signs are the same as the staggered
pattern, $\vecR$ is the position of the honeycomb unit cell,
$\veck_{\rm M}$ is the wavevector of M-point (\fig{fig:lattice}).

Results are summarized in \tbl{table:enum}. For the isotropic
Kagom\'e model, the staggered pattern mode has the largest
mean-square value, while for the distorted Kagom\'e model the
chirality stripe pattern has the largest mean-square value, which is
consistent with the ensemble-averaged spin-spin correlation result.
Also from the scaling of the mean-squares with system size we
conclude that there is no long-range-order for chirality variables
at these Fourier modes.

\begin{table}
\caption{Exact enumeration results for $L\times L$ open boundary
chirality lattice in the geometry of \fig{fig:lattice}. The number
of classical ground states $N_{\rm GS}$ for isotropic and distorted
Kagom\'e lattices are shown. The tendency to order in different
patterns [$q=0$ ($\Gamma$), $\sqrt{3}\times\sqrt{3}$ (K) and
stripe (M) patterns] are compared by evaluating mean-square values
of relevant chirality combinations
[$\av{m_\Gamma^2}$, $\av{m_{\rm K}^2}$, $\av{m_{\rm M}^2}$ respectively].}
\begin{tabular}{|r|rrr|rrr|}
\hline\hline
 & \multicolumn{3}{c|}{Kagom\'e} & \multicolumn{3}{c|}{distorted}\\
L & $N_{\rm GS}$ &  $\frac{\av{m_\Gamma^2}}{\av{m_{\rm K}^2}}$ &  $\frac{\av{m_{\rm M}^2}}{\av{m_{\rm K}^2}}$
& $N_{\rm GS}$ &  $\frac{\av{m_\Gamma^2}}{\av{m_{\rm K}^2}}$ &  $\frac{\av{m_{\rm M}^2}}{\av{m_{\rm K}^2}}$ \\
\hline
1 & 22 & 0.50 & 1.00 & 14 & 0.64 & 1.00  \\
2 & 952 & 0.32 & 0.90 & 168 & 0.62 & 1.03 \\
3 & 84,048 & 0.22 & 0.92 & 1,864 & 0.61 & 1.25 \\
4 & 15,409,216 & 0.17 & 0.84 & 19,724 & 0.61 & 1.25 \\
5 & & & & 201,584 & 0.61 & 1.31 \\
6 & & & & 2,008,276 & 0.61 & 1.35 \\
7 & & & & 19,596,536 & 0.61 & 1.45 \\
8 & & & & 188,078,644 & 0.60 & 1.41 \\
9 & & & & 1,779,795,056 & 0.60 & 1.48 \\
\hline\hline
\end{tabular}
\label{table:enum}
\end{table}

\subsection{Comparison with $^{51}$V NMR in Volborthite}
We have already noted that in contrast to classical ground states on
the isotropic Kagom\'e lattice, all ground states on the distorted
Kagom\'e lattice are disconnected from one another, and require
moving an infinite number of spins. Within a semi-classical
viewpoint, large kinetic barriers separating the distorted Kagom\'e
ground states might lead to freezing at low temperatures.
Interestingly, low temperature NMR experiments \cite{Bert:} on
Volborthite indicate spin freezing below 1.5K($\sim
J/60$) \cite{note1}, but no such freezing is observed in the
isotropic Herbertsmithite \cite{Helton:,Ofer:,Mendels:,Imai:}. It is
tempting to attribute this difference in behavior to the difference
in connectivity of classical ground states in the two cases. The
vanadium atoms occupy the hexagon centers of the Kagom\'e lattice,
and are hence coupled to six spin-1/2 $Cu$ moments.
Experimentally, on cooling through the glass transition temperature
there is a rapid rise of $1/T_1$, and at lower temperatures two
distinct local environments for the $^{51}$V sites appear, a higher
static field environment (rectangular lineshape) estimated to
involve 20$\%$ of spins, and a lower field environment (gaussian
lineshape) for the remainder. We assume that the glassy state
locally resembles {one of the classical ground states, and that
they occur with equal probability}. Then a volume average of a local
quantity in the glassy state corresponds to an ensemble average over
classical ground states. Of relevance to the NMR experiments here is
the distribution of exchange fields at the $^{51}$V site, arising
from spin configurations on the hexagons. For the nearly isotropic
case $\alpha \approx 1$, three different field values ($H$) are
possible, $H\approx3H_{cu}$, $H\approx \sqrt{3}H_{Cu}$ and
$H\approx0$, where $H_{Cu}$ is the field from a single spin. For
example, the first corresponds to a local $\sqrt{3}\times\sqrt{3}$
pattern with staggered chirality. We need to calculate the
probability to find these different fields.



The authors of \Ref{Bert:} put forward the interesting
suggestion that the high field component seen in NMR corresponds to
local $\sqrt{3}\times\sqrt{3}$ pattern. {Their arguments though
rested on properties of the isotropic Kagome model. Here, we
analytically evaluate the probability distribution of different
field configurations for the distorted Kagome lattice using the
transfer matrix method (details in Appendix A).} The probability of
obtaining the $3H_{cu}$ exchange field is found to vanish in the
thermodynamics limit, while that of the $\sqrt{3}H_{Cu}$ is $25\%$
and of the approximately zero field configuration is $75\%$. This is
roughly consistent with the experimental observation, but implies a
revised value for the local moment that was obtained in
Ref.~\onlinecite{Bert:} which assumed a local field of $3H_{Cu}$.
Hence we anticipate a copper moment per site of $0.4\times \sqrt{3}
= 0.7$ of the full moment, for small anisotropy. If the anisotropy
is significant, the local field also changes, with the previous
$\sqrt{3}H_{Cu} \rightarrow \sqrt{(5\alpha-2)/\alpha^3}H_{Cu}$ and
the zero field values now being $|2-2\alpha^{-1}|H_{Cu}$ (with 50\%
probability) and $\alpha^{-2}|\alpha-1| H_{Cu}$ (with 25\%
probability). This suggests an upper bound for the anisotropy by
requiring the local moment be less than unity, which gives
$\alpha<1.6$.

\section{Effect of Fluctuations about the Classical Ground States}
\label{sec:orderbydisorder}
It is well-known that thermal or quantum
fluctuation can lift the classical ground state
degeneracy\cite{Villain:}. In the isotropic Kagom\'e model these
kinds of `order-by-disorder' studies suggest that the Kagom\'e
antiferromagnet would select the $\sqrt{3}\times\sqrt{3}$ ground
state\cite{Chubukov:,Henley:,Chandra:,Huse_Rutenberg:,Reimers}, namely the
staggered chirality pattern.

We study the `order-by-disorder' effect in the distorted Kagom\'e
model ($\alpha>1$) by quantum and classical `spin wave' theory. It
is found that at quadratic order the fluctuations (quantum or
classical) cannot distinguish different coplanar classical ground
states. One has to go beyond quadratic order of fluctuation to find
`order-by-disorder' phenomenon.

\subsection{Linear Spin Wave Theory}
A classical coplanar ground state can be described by angles
$\theta_j$ of classical spins with respect to a reference direction
in spin space. Define a local spin axis for every site such that the
$S^z$ axis is perpendicular to the common plane of all classical
spins, and the $S^x$ axis is along the classical spin direction.

The Hamiltonian becomes
\be
\begin{split}
H=\sum_{<ij>}{}&J_{ij}[S^z_i S^z_j
+\cos(\theta_{ij})(S^x_i S^x_j+S^y_i S^y_j)\\
&-\sin(\theta_{ij})(S^x_i S^y_j-S^y_i S^x_j)]
\end{split}
\label{equ:Hrotated}
\ee
where $\theta_{ij}=\theta_i-\theta_j$ is the angle between classical spins on
sites $i$ and $j$, and the chiralities determine the sign of these angle differences.

For quantum spin-$S$ spins we can use the Holstein-Primakoff bosons to
describe the fluctuations
\be
\begin{split}
S^x_i&=S-n_i\\
S^+_i&=S^y_i+\im S^z_i=\sqrt{2S-n_i}\cdot b_i\\
S^-_i &=S^y_i-\im S^z_i=b_i\yd\sqrt{2S\nd-n_i\nd}
\end{split}
\nn\label{equ:HPboson}
\ee
where $n_i=b\yd_i b\nd_i$ is the boson number operator.

Expanding in powers of $1/S$, the Hamiltonian becomes
\be
H=E_{\rm GS}+S^{3/2}H_1+S\cdot H_2+S^{1/2}H_3+H_4+\dots
\nn\label{equ:Sexpansion}
\ee
where $E_{\rm GS}$ is the classical ground state energy, $H_n$ contains $n$-th
order boson creation(annihilation) operator polynomials.
In fact $H_1$ identically vanishes.
$H_2$ gives the quadratic (or so-called `linear') spin wave theory.
\be
\begin{split}
H_2=&\sum_{<ij>}{H_{2,ij}}\\
H_{2,ij}=&-J_{ij}\cos(\theta_{ij})[n\nd_i+n\nd_j
-(1/2)(b\yd_i+b\nd_i)(b\yd_j+b\nd_j)]\\
&-(1/2)J_{ij}(b\yd_i-b\nd_i)(b\yd_j-b\nd_j)
\end{split}
\label{equ:spinwaveH2} \ee Notice that $H_2$ only depends on
$\cos(\theta_{ij})$, then it is identical for all classical ground
state configurations ($\theta_{ij}$ can differ only by a sign between
different classical ground states). Therefore spin wave expansion
at the quadratic level cannot lift the classical degeneracy.

Dispersion of the quadratic spin wave is presented in
\app{app:spinwavedispersion}. One interesting result is that
although the dispersion becomes much more complicated than that of
the isotropic Kagom\'e model, the zero-energy flat band still
exists. Another strange feature is that as long as $\alpha\neq 1,\
\alpha > 1/2$, the `spin wave velocity' vanishes in the direction
perpendicular to the BC-bonds.

\subsection{Classical `Spin-Wave' Expansion and Effective Chirality Hamiltonian}
To lift the classical degeneracy we need to consider
the `non-linear' spin wave theory, especially the cubic order terms $H_3$,
because they are the lowest order terms distinct for
different classical ground state configurations.
Following Henley and Chan\cite{Henley:} we can in principle derive
the effective interactions between chirality variables.
In the remaining part of this section we use a different formalism by combining
Henley's idea and the classical low temperature `spin wave' expansion\cite{Brezin, Chalker, Bergman:}.

We consider classical O(3) spins on the distorted Kagom\'e lattice.
To simplify the notations we set the spin length $S$ to unity. We
define local spin axis as in previous subsection, $S^z$ axis
perpendicular to all spins, $S^x$ axis along the classical spin. We
can still use the expression \eqn{equ:Hrotated} for the Hamiltonian.
For classical spin it is convenient to parametrize the fluctuation
by \be S^y=\flucy,\,S^z=\flucz,\,S^x=\sqrt{1-(\flucy)^2-(\flucz)^2}
\nn \ee and the in-plane $\flucy$ and out-of-plane $\flucz$
fluctuations are supposed to be small at low temperatures.

The most important contributions to the partition function
comes from fluctuations around classical ground states.
\be
\begin{split}
\mathcal{Z}&=Z_0^{-1}\int{\mcD \vecS \exp(-\beta H)\prod_{i}{\delta[(\vecS_i)^2-1]}}\\
&\propto\sum_{\rm classical\ GS}{\int{\mcD \flucy \mcD \flucz \exp(-\beta H)
\prod_{i}{(1/S^x_i)}}}
\end{split}
\nn
\ee
where $\delta[(\vecS_i)^2-1]$ is the Dirac-$\delta$ function used to
ensure unit spin length, the product $\prod{(1/S^x_i)}$ is the Jacobian of
changing variables from O(3) spin to $\flucy$ and $\flucz$.
$Z_0=(2\pi)^{3\cellno}$ is chosen in such a way that
$\mathcal{Z}\rightarrow 1$ as $\beta\rightarrow 0$
($\cellno$ is the number of unit cells).

Absorb the Jacobian into the exponential and expand $S^x$ in terms
of $\flucy$ and $\flucz$, then
the exponent becomes
\be
-\beta H=-\beta (H^y_2+H^z_2+H_3+H_4-(1/2)T\sum_{i}{\xi_i}
+\dots)
\nn
\ee
where $-(1/2)T\sum_{i}{[(\flucy_i)^2+(\flucz_i)^2]}$ comes from the Jacobian,
and to simplify the notation we define $\xi_i=(\flucy_i)^2+(\flucz_i)^2$.
Then
\be
\begin{split}
H^y_2&=\sum_{<ij>}{J_{ij}\cos(\theta_{ij})\{\flucy_i\flucy_j
-(1/2)[(\flucy_i)^2+(\flucy_j)^2]\}}\\
H^z_2&=\sum_{<ij>}{J_{ij}\flucz_i\flucz_j
-(1/2)J_{ij}\cos(\theta_{ij})[(\flucz_i)^2+(\flucz_j)^2]}\\
H_3&=(1/2)\sum_{i}{ \sum_{j}{ J_{ij}\sin(\theta_{ij})\flucy_i\xi_j } }\\
H_4&=(1/8)\sum_{<ij>}{ J_{ij}\cos(\theta_{ij})(\xi_i^2+4\xi_i\xi_j+\xi_j^2) }
\end{split}
\label{equ:H2H3H4}
\ee
Again the quadratic terms are identical for all classical ground states.

We can rescale $\flucy$ and $\flucz$ by $\sqrt{\beta}$ to absorb $\beta$
into $H^y_2$ and $H^z_2$. Define
$\rfy=\sqrt{\beta}\flucy,\rfz=\sqrt{\beta}\flucz$, then
the exponent becomes
\be
-\beta H=-\tilH^y_2-\tilH^z_2-\sqrt{T}\tilH_3-T\tilH_4-O(T^2)
\nn
\ee
where $\tilH^{y,z}_2,\tilH_3$ are obtained by replacing $\flucy,\flucz$ by
$\rfy,\rfz$ in the formulas of $H^y_2,H^z_2,H_3$, respectively.
$\tilH_4$ combines the original quartic order term $H_4$ and
the lowest order term from the Jacobian,
and we have set the Boltzmann constant $k_{\rm B}=1$.
Since higher-than-quadratic order terms are controlled by temperature,
we can do a controlled perturbative expansion in powers of the small parameter $T$.

As the first approximation we may keep only $\tilH^y_2$ and $\tilH^z_2$
for very low $T$.
Solution of the quadratic theory is presented in \app{app:classicalspinwave}.
The out-of-plane fluctuation $\rfz$ has a flat zero-energy band,
which is consistent with Moessner and Chalker's mode-counting argument\cite{Moessner:}.
The in-plane fluctuation has the `Goldstone' mode at wavevector $\veck=0$.
But since this is a classical theory, the dispersion around the `Goldstone'
mode is quadratic.

\subsection{Effective Chirality Hamiltonian}
Now we can formally write down an expansion for small $T$.
Define $\mathcal{Z}_0=\int{\exp(-\tilH^y_2-\tilH^z_2)\mcD\rfy\mcD\rfz}$.
Remember that $\mathcal{Z}_0$ is the same for all classical ground states
we are perturbing.
The free energy $f$ per unit cell for fluctuations around one classical ground state is
\be
\begin{split}
f&=(1/\cellno)E_{\rm GS}-3T\ln T
-(1/\cellno)T\ln \mathcal{Z}_0\\
&\phantom{=}-(1/2)T^2\av{(\tilH_3)^2/\cellno}_0
+T^2\av{\tilH_4/\cellno}_0+O(T^3)
\end{split}
\label{equ:EffectiveFreeEnergy}
\ee
where $\cellno$ is the number of unit cells,
$\av{\mathcal{A}}_0$ means the expectation value in the quadratic theory, {\it i.e.}
$\av{\mathcal{A}}_0=\mathcal{Z}_0^{-1}\int{\mathcal{A}\cdot\exp(-\tilH^y_2-\tilH^z_2)\mcD\rfy\mcD\rfz}$.
Since $\mathcal{Z}_0$ and $\tilH_4$ are identical for all classical ground states,
difference at $T^2$ order comes from the $\av{(\tilH_3)^2/\cellno}_0$ term.
Remember that each term in $H_3$ contains a $\sin(\theta_{ij})$,
the sign of which is determined by the chirality of the triangle containing
the bond $<ij>$.
Therefore $\av{(\tilH_3)^2/\cellno}_0$ will generate effective chirality-chirality
interactions $J_{ij}\chir_i\chir_j$ for each pair of chirality variables $\chir_i$ and $\chir_j$.
Details about calculating the chirality interactions are presented in \app{app:classicalspinwave}.

There are two technical obstacles for this `order-by-disorder' analysis:
(i) The flat zero-energy band will make the two-$\rfz$ correlation
function diverge;
(ii) The `Goldstone' mode will make the two-$\rfy$ correlation function
diverge.
Both (i) and (ii) will make $\av{(\tilH_3)^2/\cellno}_0$ divergent.

To proceed we add a term $J^z\sum_{i}{(S^z_i)^2}$ in the
Hamiltonian. This can be thought as a single-ion anisotropy term
disfavoring out-of-plane fluctuation. The flat zero-energy band will
be shifted to a positive value and no longer produce divergence. We
also need to cure the divergence from the in-plane `Goldstone' mode.
But no natural interaction can do this job. Therefore we add an
artificial mass term $J^y\sum_{i}{(\flucy_i)^2}$ to the Hamiltonian,
which gives the `Goldstone' mode a small gap, or can be thought as
an infrared cutoff. Eventually we would like to take the limit
$J^z,\,J^y \rightarrow 0$.

To check consistency we first calculated the effective chirality
interactions for $\alpha=1$ Kagom\'e model. The interactions are
antiferromagnetic and seems to be short-ranged (see
\tbl{table:chircouplings} in \app{app:classicalspinwave}). Because
the nearest-neighbor chirality antiferromagnetic coupling dominates,
the staggered chirality pattern (namely the $\sqrt{3}\times\sqrt{3}$
spin configuration) is selected, which is consistent with all
previous `order-by-disorder' studies for the isotropic Kagom\'e
model. This selection is independent of $J^z$ and $J^y$ for the
range of parameters we studied.

\begin{figure}
\includegraphics{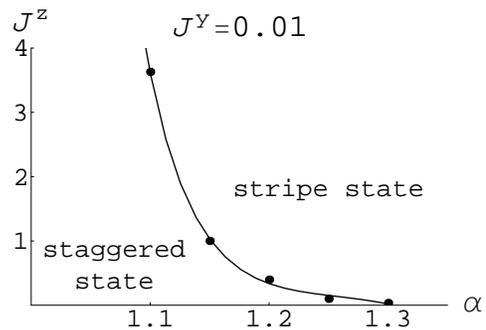}
\caption{Phase diagram obtained from classical spin-wave
`order-by-disorder' analysis with $J^y=0.01$ (artificial gap for
in-plane fluctuation). $J_z$ is the single-ion anisotropy. For large
$\alpha$ or $J^z$, the chirality stripe pattern is selected. Dots
are calculated boundary points and the line is a guide for the eye.
Weak interlayer couplings are assumed, to stabilize true long-range
order.} \label{fig:phasediagram}
\end{figure}

The $\alpha>1$ case is more delicate. It seems that the chirality
interaction is not short-ranged (see \tbl{table:chircouplings} in
\app{app:classicalspinwave}), and the selection of chirality pattern
is more sensitive to $J^z$ and $J^y$. We have calculated chirality
interactions up to sixth-neighbor, with $J^y=0.01$ as the smallest
value we can use, and for various $J^z$ and $\alpha$. A rough
picture (\fig{fig:phasediagram}) is that for $\alpha$ close to unity
or small $J_z$ the staggered chirality pattern (analogue of the
$\sqrt{3}\times\sqrt{3}$ spin configuration of the isotropic
Kagom\'e model) is still favored, but in the other part of the
parameter space our proposed chirality stripe state is selected. One
should be aware that this picture may still depend on the unphysical
parameter $J^y$, and including further neighbor chirality
interactions may also modify the phase boundary.


{
We have also tried to use this approach for $\alpha < 1$ case.
However the selection of ground state
is much more sensitive to the unphysical parameters
and the number of chirality-chirality couplings we include.
We decide to leave this part for more detailed studies in the future.
}

\section{Large-n Approximation}\label{sec:largen}
Another way to study (anti)ferromagnet is to generalize classical O(3) spin to
O($n$) spin. At $n\rightarrow \infty$ limit the theory can be solved exactly
by saddle point approximation. One can also calculate $1/n$ corrections
systematically. The saddle point approximation is supposed to be good for
high-temperature disordered phase. As temperature decreases one can usually
decide at which wavevector the long-range-order is developed, by looking at
the position of the lowest `excitation' energy, or the lowest eigenvalue(s) of
the inverse of the spin correlation function matrix.

For the isotropic Kagom\'e model the lowest `excitation' is
wavevector independent in  the saddle point solution. For distorted
Kagom\'e ($\alpha>1$) model the lowest `excitation' is degenerate on
a line in momentum space. We have to include $1/n$ correction to
determine the possible ordering wavevector uniquely. {However,
for $\alpha<1$ the $q=0$ wave-vector is selected at the saddle point
level.}

\subsection{Saddle Point Solution and Line Degeneracy for Distorted Kagom\'e Lattice}
The model we use is the O($n$) spin antiferromagnet on the distorted
Kagom\'e lattice. \be H=\sum_{\rm triangles}{\sum_{a}{ (S^a_{\rm
A}S^a_{\rm B} +\alpha\cdot S^a_{\rm B}S^a_{\rm C}+S^a_{\rm
C}S^a_{\rm A}) }} \label{equ:HOn} \ee with constraints
$\sum_{a=1}^{n}{(S^a_i)^2}=1$. We rescale all spins and $\beta$ by
$\tilS^a_i=\sqrt{n}S^a_i,\ \tilb=\beta/n$. The partition function
becomes \be \mathcal{Z}=Z_0^{-1}\int{\left (\prod_{i,\,a}{\dif
\tilS^a_i}\right ) \exp(-\tilb \tilH)\prod_{i}{\delta\left
[n-\sum_{a}{(\tilS^a_i)^2}\right ]} } \nn \ee where
$Z_0=[n^{n/2}\pi^{n/2}/\Gamma(n/2)]^{3\cellno}$ such that
$\mathcal{Z}\rightarrow 1$ as $\beta\rightarrow 0$, $\cellno$ is the
number of unit cells, $\tilH$ is the Hamiltonian $H$ with $S$
directly replaced by $\tilS$. In the remainder of this section we
will write $\tilS,\,\tilH\,{\rm and\,}\tilb$ as $S,\,H\,{\rm and
\,}\beta$, respectively. We will write $\mcD \vecS$ instead of
$\prod_{i,\,a}{\dif S^a_i}$.

Using the fact that \be \delta(x)=\int_{-\infty}^{\infty}{
\frac{\dif \lgi_i}{2\pi} \exp\left [(\im \lgi_i+\lgr_i)x\right ] }
\nn \ee where $\lgi_i$ is a real dummy variable, and $\lgr_i$ is an
arbitrary real parameter to be determined later by the saddle point
condition, we can rewrite the partition function as \be
\begin{split}
\mathcal{Z}=&Z_0^{-1}\int{\mcD \vecS \mcD \lgi
\exp\left \{-\beta H+\sum_{i}{\tillg_i\left [n-\sum_{a}{(S^a_i)^2}\right ]}\right \}
}\\
=&Z_0^{-1}\int{\mcD \vecS \mcD \lgi
\exp\left [-\sum_{a,\,i,\,j}{S^a_i M_{ij} S^a_j}+\sum_{i}{n\tillg_i}\right ]
}
\end{split}
\label{equ:OnAction}
\ee
in which $M_{ij}=(\im\lgi_i+\lgr_i)\delta_{ij}+\beta J_{ij}/2$ is
a symmetric matrix, $\tillg=\im\lgi+\lgr$,
and $\mcD \lgi=\prod_{i}{[\dif \lgi_i/(2\pi)]}$.
Integration over $S^a_i$ gives
\be
\mathcal{Z}=Z_0^{-1}\pi^{3n\cellno/2}\int{\mcD \lgi\,
\det(M)^{-n/2}\exp[\sum_{i}{n(\im \lgi_i+\lgr_i)}]
}
\nn
\ee
Now the saddle point condition is
\be
\frac{\partial}{\partial \lgr_i} \ln \det(M)=2,\quad \forall i
\nn
\ee
Let us assume the saddle point solution has all lattice symmetries, {\it e.g.}
translational invariance. Then $\lgr_i$ depends only on which sublattice
the site $i$ belongs to. Furthermore, because the B- and C-sublattices are
equivalent, we have $\lgr_{\rm B}=\lgr_{\rm C}$.

Assuming translationally invariant $\lgr_i$, the matrix $M_{ij}$ can be
block-diagonalized by Fourier transformation.
$\det(M)=\prod_{\veck}\det[M(\veck)]$ where
$M(\veck)$ is a $3\times 3$ matrix
\be
M(\veck)=\begin{pmatrix}\lgr_{\rm A} & \beta\cos(k_3/2) & \beta\cos(k_2/2)\\
\beta\cos(k_3/2) & \lgr_{\rm B} & \alpha\beta\cos(k_1/2)\\
\beta\cos(k_2/2) & \alpha\beta\cos(k_1/2) & \lgr_{\rm C}
\end{pmatrix}
\nn
\ee
with $k_i=\veck\cdot \vece_i$ ($k_3=-k_1-k_2$).
The saddle point condition becomes
\be
(1/\cellno)\sum_{\veck}{\frac{\partial}{\partial \lgr_X}
\ln \det[M(\veck)]}=2,\quad X={\rm A,B,C}
\nn
\ee
and in the thermodynamic limit $\cellno\rightarrow \infty$
the sum becomes a integral over Brillouin zone,
$(1/\cellno)\sum_{\veck}\rightarrow \int_{0}^{2\pi}{\int_{0}^{2\pi}{\dif k_1\dif k_2/(2\pi)^2}}$.

This saddle point equation cannot be solved analytically. But when $\beta$ is
small, we can expand it in terms of $\beta$ and obtain a high-temperature series
for $\lgr_X$. The result is
\begin{subequations}
\begin{eqnarray}
2\lgr_{\rm A}&=&1+4\beta^2-4\alpha \beta^3+\dots\\
2\lgr_{\rm B,C}&=&1+2(\alpha^2+1)\beta^2-4\alpha\beta^3+\dots
\end{eqnarray}
\end{subequations}
We notice that $\lgr_{\rm B,C}>\lgr_{\rm A}$ for $\alpha>1$, which
leads to a degenerate line of lowest excitation in the saddle point
approximation. This high-temperature (small $\beta$) series can be
extended to intermediate temperature ($\beta$) by Pade
approximation.

\begin{figure}
\includegraphics{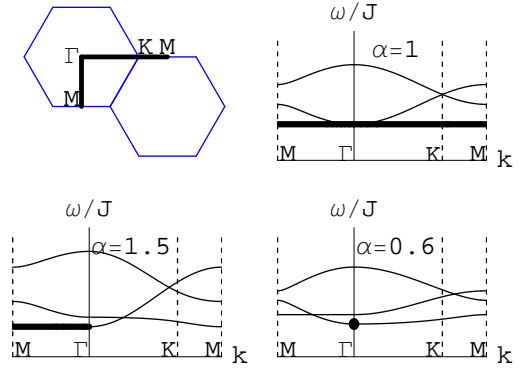}
\caption{(Color online) dispersion $\omega$ of the O($n$) model in the saddle point approximation
along certain high symmetry directions (shown in the first panel),
for three different $\alpha$, at $\beta=0.2$.
The lowest eigenvalue(s) are shown with bold lines(dot).}
\label{fig:saddlepointdispersion}.
\end{figure}

After solving $\lgr_X$ we can solve the `dispersion', or the
eigenvalues of $M(\veck)$. Dispersion along certain high symmetry
directions are shown in \fig{fig:saddlepointdispersion}. Note, for
$\alpha<1$ the lowest eigenvalue is uniquely determined at
$\veck=0$. However, for $\alpha>1$ the lowest eigenvalue is
degenerate on the $k_1=0$ line, or the vertical $\Gamma-{\rm M}$
line in the BZ. Finally, for $\alpha=1$ the lowest eigenvalue is
degenerate over the entire BZ.

\begin{figure}
\includegraphics{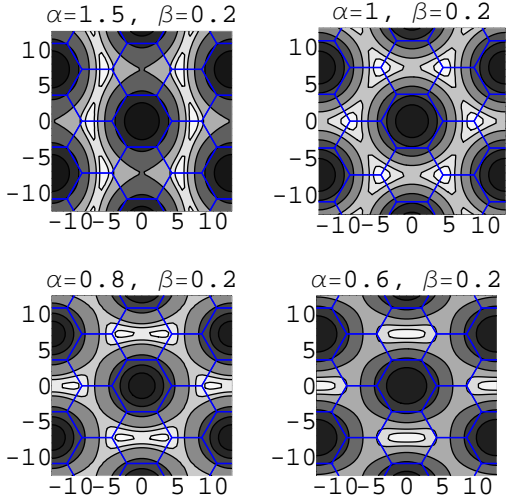}
\caption{(Color online) predicted elastic neutron scattering intensity for
distorted Kagom\'e model, obtained from the saddle point solutions
at $\beta=0.2$ for four different $\alpha$. Hexagons are BZ borders.
Darker region have lower intensities. The $\alpha>1$ case shows
quasi-1D feature.} \label{fig:OnNeutron}
\end{figure}

To decide the ordering wave-vector uniquely we must consider $1/n$
correction for $\alpha\geq 1$ cases. Before presenting that in the
next subsection, we show the calculated elastic neutron scattering
intensity (\fig{fig:OnNeutron}) $[\sum_{X,Y}{(M^{-1})_{XY}}]^2$
of the saddle point solutions for four different $\alpha$ with
relatively high temperature $\beta=0.2$ (summation is over $X,Y={\rm
A,B,C}$). We emphasize that the maximum appearing in the elastic
neutron scattering intensity does not directly correspond to the
possible long-range-order wavevector.

\subsection{Lifting the Line Degeneracy of $\alpha>1$: $1/n$ correction}
To lift the degeneracy of the lowest `excitations' of the saddle point
approximation, we have to include fluctuations around the saddle point.

We have three $\lgi_{X,\veck}$ fields and $3n$ $S^a_{X,\veck}$ fields in
the action, where $X$ is the sublattice index, $a$ is the O($n$) index of spin.
The Green's function of the spins with the same O($n$) indices is
a $3\times 3$ matrix. Under the  saddle point approximation it is
$G^{(0)}_{S,ab,XY}(\veck)=[M(\veck)]^{-1}_{XY}\delta_{ab}$
where $X,Y={\rm A,B,C}$ for three sublattices, $a,b$ are O($n$) indices.
We need the correction to this Green's function by
the fluctuations of $\lgi$ around zero. From \eqn{equ:OnAction} we see that
there is a three-leg vertex between $\lgi$ and $S^a$, of the form
$-\im\lgi_i(S^a_i)^2$.

\begin{figure}
\includegraphics{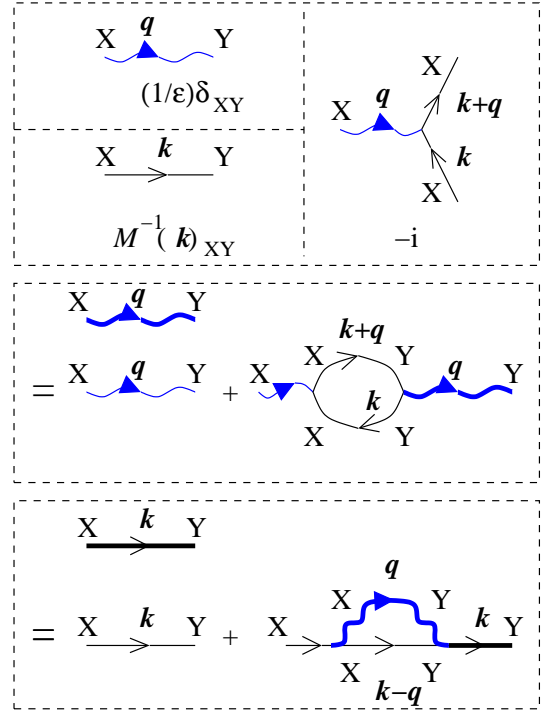}
\caption{(Color online) Feynman rules for calculating $1/n$ corrections of O($n$)
model. O($n$) indices are omitted. $X,Y$ are sublattice indices. The
first panel contains free propagators and the only vertex in the
theory. Straight lines represent the spin propagator. The second
panel is the one-loop Dyson equation \eqn{equ:Dyson1} for the $\lgi$
propagator. The third panel is the one-loop Dyson equation
\eqn{equ:DysonS} for the spin propagator. Thick lines are full
propagators.} \label{fig:OnFeynmanRules}
\end{figure}

The Feynman rules and Dyson equations are summarized in
\fig{fig:OnFeynmanRules}. Notice that the three-leg vertex preserves
sublattice index for all fields and also O($n$) index for the spins.
There is no free propagator for $\lgi$ fields in the original theory.
To make the perturbative expansion well defined we add
a term $+\sum_{i}{\epsilon(\lgi_i)^2}$ to the Hamiltonian,
which corresponds to a free
propagator $(1/\epsilon)\delta_{XY}$. Finally we will take the
$\epsilon\rightarrow 0$ limit.

The one-loop Dyson equation for the propagator of $\lgi$
is shown in \fig{fig:OnFeynmanRules}.
The inverse of the Green's function at one-loop level is
\be
\begin{split}
[G^{-1}_{\lgi}(\veck)]_{XY}=&(\epsilon)\delta_{XY}-\Gamma_{\lgi,XY}=-\Gamma_{\lgi,XY}
\end{split}
\label{equ:Dyson1}
\ee
where $\Gamma_{\lgi}$ is the self-energy of $\lgi$ (the loop diagram in
the second panel of \fig{fig:OnFeynmanRules}).
Here we have taken the $\epsilon\rightarrow 0$ limit.
\be
\begin{split}
\Gamma_{\lgi,XY}=&\sum_{a}{(-\im)^2
\int_{\rm BZ}{\dif^2 \vecq G^{(0)}_{S,aa,XY}(\veck+\vecq) G^{(0)}_{S,aa,YX}(\vecq)
} }\\
=&-n\int_{\rm BZ}{\dif^2 \vecq [M^{-1}(\veck+\vecq)]_{XY} [M^{-1}(\vecq)]_{YX} }
\end{split}
\nn
\ee
There is no summation over sublattice indices $X,Y$ on the right-hand side.
$\int_{\rm BZ}{\dif^2\vecq}$ is the normalized integral over the entire BZ.
Since the summation over O($n$) index $a$ becomes a factor of $n$,
the one-loop $\lgi$ propagator is of the order $1/n$.

We use this one-loop $\lgi$ propagator to calculate the one-loop
correction to the spin propagator.
\be
\begin{split}
[G^{-1}_{S,aa}(\veck)]_{XY}=&[M(\veck)]_{XY}-\Gamma_{S,aa,XY}
\end{split}
\label{equ:DysonS}
\ee
where $\Gamma_{S,XY}$ is the self-energy of spins (the loop diagram in
the third panel of \fig{fig:OnFeynmanRules}).
\be
\begin{split}
\Gamma_{S,aa,XY}=&(-\im)^2\int_{\rm BZ}
{\dif^2 \vecq G^{(0)}_{S,aa,XY}(\veck-\vecq) G_{\lgi,XY}(\vecq) }
\end{split}
\nn
\ee
Again there is no summation over $X,Y$ on the right-hand side.

These integrals cannot be evaluated exactly. Instead we use the
high-temperature (small $\beta$) expansion to get analytical result.
We found that up to $\beta^7$ order the one-loop correction does not
qualitatively change the form of $G^{-1}_{S,aa}(\veck)$. It has
similar wavevector dependence of the inverse free propagator $M(\veck)$,
therefore the line degeneracy of $\alpha>1$ model and the degenerate band
of $\alpha=1$ model cannot be lifted at $\beta^7$ order.

However at $\beta^8$ order a qualitatively distinct correction appears.
The self-energy (the loop diagram) contains a term
\be
(1/n)\beta^8 C
\begin{pmatrix}
0 & \alpha\cos(\frac{k_{12}}{2}) & \alpha\cos(\frac{k_{13}}{2})\\
\alpha\cos(\frac{k_{21}}{2}) & 0 & \frac{\lgr_{\rm BC}}{\lgr_{\rm A}}\cos(\frac{k_{23}}{2})\\
\alpha\cos(\frac{k_{31}}{2}) & \frac{\lgr_{\rm BC}}{\lgr_{\rm A}}\cos(\frac{k_{32}}{2}) & 0
\end{pmatrix}
\nn
\ee
where $k_{ij}=k_i-k_j$ and a constant $C=\alpha^2/(64\lgr_{\rm A}^2\lgr_{\rm BC}^5)$.
This looks like a next-neighbor ferromagnetic coupling.

For $\alpha>1$ model we have a degenerate line $k_1=0$ at the saddle point level.
This $(1/n)\beta^8$ correction will favor $k_2=\pi$ which is the M-point in
the BZ.
For $\alpha=1$ model we have a degenerate band in the saddle point
approximation.
This $(1/n)\beta^8$ correction will favor $k_1=k_2=2\pi/3$ which is the
K-point in the BZ, corresponding to the $\sqrt{3}\times\sqrt{3}$ spin
configuration.

We notice that a previous high temperature series expansion
study\cite{Harris:} also lifts the degeneracy of the Kagom\'e O($n$)
model at $\beta^8$ order. Their result contains, in some sense,
corrections to all orders of $1/n$, but do not have a simple
analytical form. Our simpler analytic method (expanding in both
$1/n$ and $\beta$) is complementary to their linked-cluster series
expansion study and our results are consistent with theirs in the
region of overlap.

\section{Quantum Limit: Exact Diagonalization and Slave Particle Mean Field
Theories}\label{sec:quantumstudy} We have performed exact
diagonalization on small lattices of spin-1/2 moments, to study the
effect of distortions in the Kagome model. We used the open source
ALPS library and applications\cite{ALPS} on an office computer. Two
different kinds of results are presented; the nature of the ground
state and low lying excitations, and the thermodynamics (specific
heat and magnetic susceptibility). The latter requires knowing all
eigenvalues of the Hamiltonian, and is hence restricted to small
system sizes of 12 sites ($2\times 2$ unit cells) with periodic
boundary conditions. Based on previous studies\cite{Elstner:} we
believe that this small system can still produce qualitatively
correct high temperature properties. For the former, we study system
sizes upto 24 sites ($4\times 2$).

\begin{figure}
\includegraphics{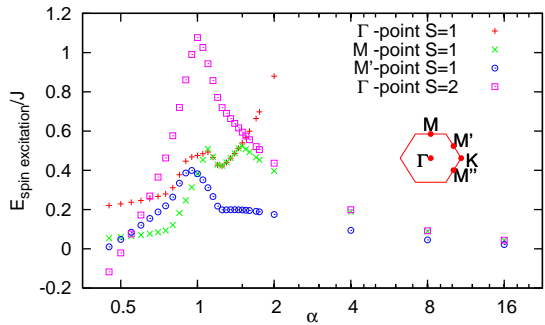}
\caption{(Color online) spin excitation energies at different
wavevectors from the exact diagonalization study (12 sites: $2\times
2$ unit cell system with periodic boundary condition). Energy is
measured from the lowest singlet state (ground state for $\alpha > 1/2$). 
Other spin excitations, $S > 2$ at any wavevector or $S = 2$
away from the $\Gamma$-point, have much high energies than the ones
plotted. } \label{fig:gaps}
\end{figure}

Before discussing the results of exact diagonalization for the low
lying eigenstates, let us briefly recall the expectation from the
semiclassical picture developed so far.
\begin{enumerate}
\item For $\alpha < 1/2$ a colinear ferrimagnetic state with a
magnetization of 1/2 per unit cell is expected.
\item For $\alpha>1$, the ferrimagnetic chirality stripe state is
expected, which implies a net spin in the ground state and low energy
spin excitations at the $M$ point as shown in the inset of \fig{fig:gaps}.
\item For $1/2 < \alpha < 1$, we do not have a firm expectation from
semiclassics, however, the large-$n$ saddle point solution seems to
favor a $q=0$ state, which would also be ferrimagnetic.
\end{enumerate}

Numerically, we find that the first prediction is remarkably well
obeyed even in this extreme quantum limit. On decreasing $\alpha$,
the ground state is found to have non-zero total spin. Moreover,
this is found to happen precisely below $\alpha = 1/2$. The ground
state moment is also exactly what is expected, for example it is
$S=2$ for the $2\times 2$ lattice and $S=3$ for the $3\times 2$
lattice. For $\alpha > 1/2$ the comparison is less clear. For example,
the ground state is a spin singlet on lattice sizes up to 24 sites.
However, there is a clear tendency of the $S=2$ state at the
$\Gamma$ point to drop in energy on moving away from the isotropic
Kagome point as seen in \fig{fig:gaps}, 
indicating perhaps a
tendency to develop a net moment. On the other hand, while the spin gap
may be expected to be soft along the $M$ point (the wavevector for the
chirality state) for $\alpha>1$, it turns out that the $M$ point
is actually not the location of the lowest spin carrying excitation - 
which instead occurs at different wavevectors; 
the $M'$ (and equivalent $M''$) locations in
the case of $2 \times 2$ system. Similarly, for the $4\times 2$
system, the $S=1$ excitation energy at the $M$ point is higher than
those at the $M'$ and $M''$ points (the latter two are inequivalent on
this lattice). Moreover the lowest $S=1$ excitation occurs at the
$M''$ point.

\begin{figure}
\includegraphics{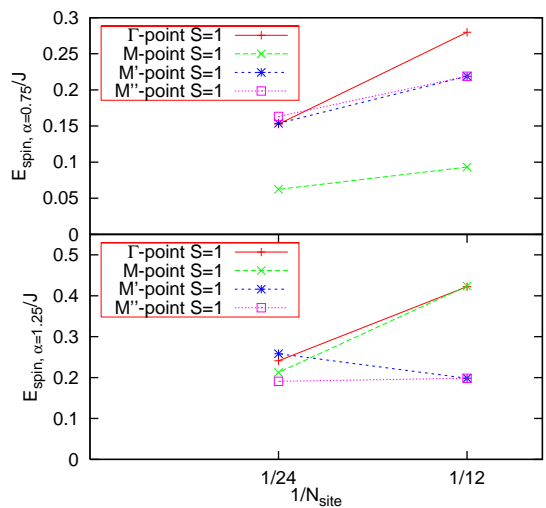}
\caption{(Color online) spin excitation energies at the high
symmetry points (see inset in \fig{fig:gaps} for definition) vs.
number of sites, for $\alpha=0.75$ and 1.25, for 12 site and 24 site
systems. Due to the geometry of the 24-sites ($4\times 2$ unit cell)
lattice, the M' and M'' points are not equivalent.
Lines are guides for the eye.}
\label{fig:gapvsN}
\end{figure}

It should be noted though that the $S=1$ excitation
energy at the $M$ point decreases rapidly from the 12 site to the 24 site lattice
\ref{fig:gapvsN}, and might end up being the lowest spin excitation at
larger system sizes. Paradoxically, in the $1/2 < \alpha < 1$ limit, the
$M$ point is the location of the lowest spin carrying excitation, both
in the 12 and 24 sized systems we studied. 
We therefore have to leave open the question of the
validity of the semiclassical `chirality stripe' picture in the
extreme quantum limit, to future systematic numerical studies on
bigger systems. Finally, we note that as in the isotropic kagome
case, we have observed singlet excitations inside the spin-gap
(the energy of the lowest excitation with non-zero spin).

\begin{figure}
\includegraphics{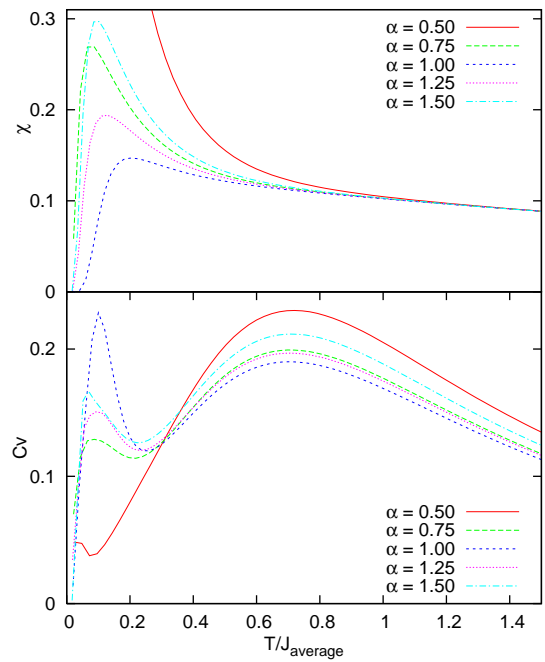}
\caption{(Color online) susceptibility $\chi$ and specific heat Cv from the exact
diagonalization study ($2\times 2$ unit cell system, 12 spins,
with periodic boundary condition).
Temperature is rescaled by the average coupling $J_{\rm average} =
(2+\alpha)/3$ for each curve. The susceptibilities of different
$\alpha>0.5$ converge to the $\alpha=1$ result even at moderate
temperatures. The positions of specific heat maxima at around
$T/J_{\rm average}=2/3$ are consistent  between different $\alpha$
values. } \label{fig:EDchiCv}
\end{figure}

{\em Thermodynamics:} The magnetic dc-susceptibility and specific
heat results for several different $\alpha$ are presented in
\fig{fig:EDchiCv}. In both figures the temperature has been rescaled
by the average coupling $J_{\rm average}=(2+\alpha)/3$ for each
curve and $\chi$ is also rescaled accordingly. For high-temperature
($T>0.2J_{\rm average}$) the dc-susceptibilities for different
$\alpha$ converge to the $\alpha=1$ result. The positions of the
broad maxima in the specific heat curves are also more or less the
same for different $\alpha$. Therefore we conclude that the
anisotropy does not induce qualitative difference in these two
macroscopic observables for high enough temperature ({\it e.g.}
$T>0.2\,J_{\rm average}$ ).

{\em Slave Particle Approaches:} Other theoretical approaches can
also be used to attack the problem directly from the quantum limit.
These methods have been applied to the isotropic Kagom\'e lattice
and can be utilized to study the effect of distortion. The Schwinger
boson technique (large-N Sp(N) approach) has been used to study the
Volborthite lattice recently \cite{Apel:}, where for not too large
spatial anisotropy the $\sqrt{3}\times\sqrt{3}$ state was found to
persist, although the ordering wavevector is shifted to an
incommensurate value (the staggered chirality pattern remains the
same). Fermionic slave particle representation of the
spins\cite{RanY:} as well as the dual vortex
formulation\cite{Alicea:} have recently been used to study the
isotropic Kagom\'e lattice in connection to Herbertsmithite.
Extending these studies to the Volborthite lattice should be
interesting. For example, the Dirac fermions in the proposal of
Ref.~\onlinecite{RanY:} would remain massless on the distorted
lattice as well, since the mass term is prohibited by the
translational and time reversal symmetries that remain intact.


\section{Conclusions}
We have studied the distorted Kagom\'e model by several approaches.
First we proved that the classical degeneracy is reduced from an
extensive one (of the isotropic Kagom\'e model) to a sub-extensive
one. As a result, we found that the ground state ensemble is much
less connected in the distorted Kagom\'e model compared to the
isotropic case. One has to change an infinite number of spins (in
the thermodynamic limit) in order to move from one classical
coplanar ground state to another. This could result in a greater
tendency towards glassy behavior and may be consistent with the fact
that spin freezing was observed (not observed) in Volborthite
(Herbertsmithite). We then studied the properties of the ground
state ensemble by enumeration and transfer matrix methods. Using
transfer matrix method we calculated the probability of different
local spin configurations and showed that this consideration may
provide an explanation of the low temperature NMR data in
Volborthite.

We then studied how this remaining degeneracy can be lifted by two
novel refinements of various approaches to the classical problem. In
particular, we used a low temperature classical spin-wave expansion
to compute the effective chirality interactions which lead to a
preferred ordering pattern. We also studied the large-$n$ O($n$)
model in the saddle point approximation and with $1/n$ corrections,
the latter performed in conjunction with a high temperature
expansion. Our results for the isotropic case $\alpha=1$ are
consistent with previous order-by-disorder studies for the isotropic
Kagom\'e model, {\it i.e.} $\sqrt{3} \times \sqrt{3}$ state is
selected. However for $\alpha>1$, both classical approaches we
pursued point to a possible long-range-order pattern different from
that of the isotropic Kagom\'e model. The resulting `chirality
stripe state' doubles the magnetic unit cell, has a Fourier
component at the M-point in the Brillouin zone, and has a net
magnetic moment (\fig{fig:lattice}). Of course, this classical 2D
system cannot develop a long-range-order at any finite temperature,
but in the presence of weak inter-layer couplings, the ordering
pattern we propose is the most reasonable candidate if magnetic
long-range-order sets in. Exact diagonalization studies of small
systems showed that the specific heat and susceptibility for
different values of $\alpha$ do not vary much at intermediate
temperatures upon the change of the anisotropy parameter $\alpha$.

Shortly after completion of this work there appeared another
paper\cite{Yavorskii} studying the same lattice but via the Sp(N)
large-N treatment and perturbation theory. Their analysis of the
degree of classical degeneracy is in agreement with our result.

\section{Acknowledgements}
We thank Doron Bergmann, Leon Balents, and John Hopkinson for
useful discussion. We acknowledge
support from the Hellman Family Faculty fund, LBNL DOE-504108 (F.
W. and A.V.), the NSERC of Canada, Canadian Institute of Advanced Research,
Canada Research Chair Program, KRF-2005-070-C00044, and
Visiting Miller Professorship at University of California at Berkeley (Y.B.K.).
Some part of this work was done at the Kavli Institute for Theoretical Physics
at University of California at Santa Barbara and is supported in part by
the NSF Grant No. PHY05-51164.

\appendix

\section{Transfer matrix solution of the classical ground state degeneracy of
the distorted Kagom\'e model}\label{app:Tmatrix} In this appendix we
derive the asymptotic formula of the classical ground state
degeneracy in the distorted Kagom\'e model, and also establish
rigorous upper and lower bounds to show that the degeneracy is
sub-extensive. We also study the probability of various local
hexagon configurations in the ground state ensemble of the distorted
Kagom\'e model, which is related to NMR studies of the
Volborthite\cite{Bert:}.

\begin{figure}
\includegraphics{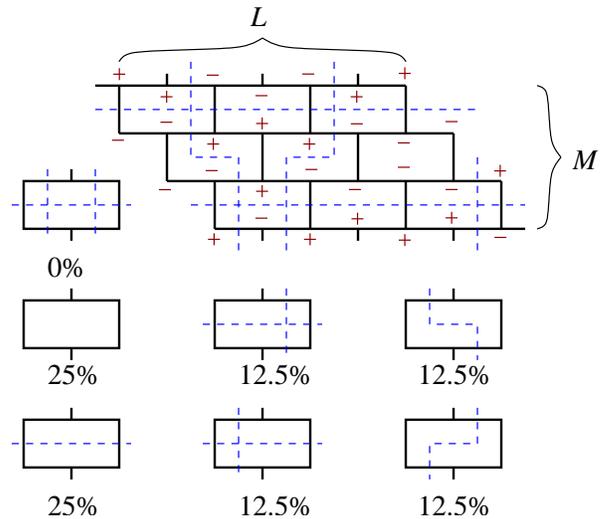}
\caption{(Color online) brickwall lattice for the transfer matrix study in \app{app:Tmatrix}.
Seven possible single `brick' configurations and their probabilities in
the thermodynamic limit are presented.}
\label{fig:brickwall}
\end{figure}

We stretch the honeycomb chirality lattice horizontally
to make a topologically equivalent `brickwall' lattice (\fig{fig:brickwall}).
Chiralities are Ising variables on the vertices.
For simplicity of derivation we use a different, less symmetric,
geometry other than the geometry used for enumeration study in the main text.
The lattice consists of $M$ rows of `bricks', each row contains $L$ `bricks'.
We will establish the upper and lower bounds, $4^{M+L}$ and $2^{M+1}$, for
open boundary condition, and the asymptotic formula $2^{M+L}$ for periodic
boundary condition in the thermodynamic limit.

It is better to represent the states of the Ising chirality variables by
domain wall configurations (\fig{fig:brickwall}).
As in all Ising systems, the number of Ising configurations is
two times the number of domain wall configurations.
All possible domain wall configurations within a `brick' is given in
\fig{fig:brickwall}. Number below each `brick' configuration is
the probability of that local configuration in a lattice with periodic
boundary condition in the thermodynamic limit, to be derived later.

There are several important observations:\\
(a) If there is a horizontal domain wall crossing one of the vertical edges
of the `brickwall', this domain wall must extend through the entire lattice,
cutting the entire row of `bricks'.\\
(b) The number of vertical domain walls cutting a horizontal line
in the `brickwall' is conserved from line to line.\\
(c) Whether there is a horizontal extended domain wall in the row of `bricks'
or not completely determines the propagation of vertical domain walls from the
upper line to the lower line.\\
(d) If there are two vertical domain walls in the same `brick' in
the upper line (we call this a `collision' of two vertical domain walls),
then there must be a horizontal extended domain wall
in the row of `bricks', and we have only one choice for the vertical
domain wall configuration on the lower line.
Otherwise for a given vertical domain wall configuration on the upper line
we have two choices on the lower line.\\
(e) Vertical domain walls do not cross each other.

We can obtain an upper bound for the number of chirality configurations by the following
considerations for a lattice with open boundary condition.\\
(i) The vertical domain wall configurations on the topmost line give $2^{2L}$ choices;
(ii) The horizontal extended domain walls give a factor of {\em at most} $2^{M}$;
(iii) On each row of `bricks' except for the first row there {\em could} be
one additional Ising degree of freedom depending on whether there is
a vertical domain wall entering from the top-right edge of the rightmost `brick'
(an example of entering vertical domain wall is given in \fig{fig:brickwall}
- the third row from top).
This is {\em at most} a factor of $2^{M-1}$.
Combining all these factors we get an upper bound $4^{M+L}$ for chirality
configurations on the $L\times M$ open boundary lattice.

We can easily get a sub-extensive lower bound for open boundary condition
by considering the case that there is no vertical domain wall. Then we have $2^{M}$
domain wall configurations via the $M$ possible horizontal extended domain walls.
Thus a lower bound of the number of chirality configurations is $2^{M+1}$.

Now we impose the periodic boundary condition on an $L\times M$ `brickwall'.
Strictly speaking the periodic boundary condition will introduce two additional
non-local constraints on the chirality variables. And it will impose
constraints on the total number of vertical domain walls (must be even)
and also horizontal domain walls. They are not supposed to change
the asymptotic behavior and we ignore them for simplicity.

Define the transfer matrix $T_{xy}$, where $x,y$ label the vertical
domain wall configurations on the upper and lower line of a row of
`bricks', respectively. $T_{xy}$ is the number of ways that vertical
domain walls in $x$ can propagate downward to $y$. Some examples:
(i) $x$ is the configuration where there is no vertical domain wall
in a line, then the only $y$ satisfying $T_{xy}\neq 0$ is $y=x={\rm
(no\ vertical\ domain\ wall)}$, and $T_{xx}=2$ because there could
be one, or no, extended horizontal domain wall in between, which
should be counted as two different ways of propagation; (ii) $x$ is
the configuration where there are vertical domain walls on every edge
of the upper horizontal line, then $T_{xx}=1$ because there must be
one extended horizontal domain wall in between, and $T_{xy}=0,\
\forall y\neq x$.

The number of domain wall configurations is the trace of the $M$-th power of
the $2^{2L}\times 2^{2L}$ transfer matrix $T$, which equals the sum of the
$M$-th powers of all eigenvalues $\lambda$ of $T$,
${\rm Tr}(T^{M})=\sum_{\lambda}{\lambda^{M}}$.
From the previous observations (c) and (d) we have $\sum_{y}{T_{xy}}\leq 2$.
Therefore all eigenvalues have absolute values smaller than or equal to 2.
This provides an upper bound $2^{2L+M}$ for domain wall configurations.

Take the thermodynamic limit $M\rightarrow \infty$, with $L$ large but finite,
then the trace ${\rm Tr}(T^{M})$ reduces to the sum of the $M$-th power of
the largest eigenvalues (it is 2 and can be degenerate),
$\sum_{\lambda=2}{2^{M}}$. Now we want to construct all eigenvectors
corresponding to eigenvalue 2.
The property of the ground state ensemble is dominated by these eigenvectors
in the thermodynamic limit.

Suppose $a_{x}$ is a (left) eigenvector with the eigenvalue 2,
$\sum_{x}{a_{x}T_{xy}}=2a_{y}$. Then we have the following two properties:
(i) $a_{x}\geq 0,\ \forall x$, this comes from the fact that $T_{xy}\geq 0$;
(ii) $a_{x}=0$ for $x$ containing a `collision', this comes from the observation
(d).

If there is no `collision' in $x$, but
there is one vertical domain wall crossing the top-left horizontal edge of
one `brick', and one of its neighboring vertical domain wall crosses
the top-right horizontal edge of another `brick',
we can always bring those two vertical domain walls together to make
a `collision', by propagating them downward
(an example is shown in \fig{fig:brickwall}).
Therefore we must have $a_{x}=0$ for this kind of $x$,
which contains both vertical domain walls crossing top-left and
top-right edges of some `bricks'.

Now we can construct all eigenvectors with the largest eigenvalue(2).
Take an $x$ containing vertical domain walls crossing only
the top-left edges of some `bricks'.
It can propagate to the next line without change,
or shifted by one half of the lattice constant.
By translating this $x$ on the line
(with periodic boundary condition) by multiples of half lattice constant,
we find a connected subspace of the state space,
denoted by ${\rm span}(x)$. Then $a_{y}=1,\ \forall y\in{\rm span}(x)$ is the
(not normalized) eigenvector with the largest eigenvalue(2) in this subspace
(by Perron-Frobenius theorem this eigenvector is unique in this subspace).

The degeneracy of the largest eigenvalues(2) equals to the number
of distinct subspaces constructed as in the previous paragraph, or the
number of inequivalent $x$ with only the top-left-edge vertical domain
walls (inequivalent under translation). This is still a non-trivial
combinatorial problem, but we have a rough upper bound $2^L$ and a
lower bound $2^L/L$. Combining all previous considerations we have
the asymptotic form of the number of configurations $2^{M+L}$.

Now we have, in principle, all the eigenvectors relevant in the
thermodynamic limit. We can find the probabilities of every `brick'
configuration, or the configuration of the six spins in a hexagon in
the original distorted Kagom\'e lattice. This is related to the
$^{51}$V NMR study in Bert {\it et al.}\cite{Bert:}, because
different local spin configurations will produce different magnetic
field on the V site. However the authors of that experimental paper
did not take into account the constraints on chirality variables,
thus their theoretical estimates of the probabilities of different
local configurations are incorrect.

First we consider the `brick' configuration containing a `collision' of
vertical domain walls. This corresponds to the local $\sqrt{3}\times\sqrt{3}$
configuration, which produces the largest magnetic field
(3 times of a single Cu if $\alpha\sim 1$,
in general the factor is $2+2\alpha^{-1}-\alpha^{-2}$)
on the V site. However since our eigenvectors do not contain `collision'
the probability of this local configuration is zero.

Next we consider the configuration where there is one vertical domain wall
and also one horizontal domain wall through the `brick'. This will produce a
smaller magnetic field ($\sqrt{3}$ times of a single Cu if $\alpha\sim 1$,
in general the factor is $\sqrt{(5\alpha-2)/\alpha^3}$).

Notice that we have a particle-hole like symmetry.
For a subspace ${\rm span}(x)$ discussed in the previous paragraphs,
where $x$ contains vertical domain walls through some of the top-left
edges of `bricks',
we can construct another subspace ${\rm span}(\bar{x})$ from a `complementary'
configuration $\bar{x}$, in which there is one vertical domain wall through
a top-left edge of a `brick' if and only if
there is no vertical domain wall through that edge in $x$.

Therefore the probability that there is one vertical domain wall through
the `brick' is one half.
The probability of a horizontal domain wall through the `brick' is clearly
also one half for the eigenvectors we consider.
Combining these two factors we have the probability 25\%
for this type of local configuration.
Note that whether the vertical domain wall is on the left- or right-side
will give another factor of one half,
hence the 12.5\% probabilities in \fig{fig:brickwall} for the
two configurations of this type.

Probability of other configurations can be derived in the similar fashion.
But all the other local configurations will produce very small
magnetic field on the V site (for $\alpha\sim 1$).
In particular, the configuration with no domain wall through the `brick'
has a magnetic field $|2-2\alpha^{-1}|$ times a single Cu field,
with the probability 25\%.
The configuration with no vertical domain wall but a horizontal domain
wall has the same magnetic field factor $|2-2\alpha^{-1}|$,
with the probability 25\%.
The two configurations with one vertical domain wall but no horizontal
domain wall have the magnetic field factor $|\alpha-1|/\alpha^2$,
and the total probability 25\% (12.5\% each).

Based on these analyses we argue that the 20\% slow component
observed in NMR\cite{Bert:} is not due to the local
$\sqrt{3}\times\sqrt{3}$ configuration, but rather the
configurations producing a smaller (factor $\sqrt{3}$ rather than 3)
magnetic field and with a theoretical probability 25\% (with one
vertical and one horizontal domain wall).

\section{Dispersion of quadratic quantum spin wave}\label{app:spinwavedispersion}
In this appendix we present the quadratic (or the so-called
`linear') quantum spin wave dispersion of the distorted Kagom\'e
Heisenberg model. We notice that there is still a zero-energy band,
and the `spin wave velocity' of the dispersive branch vanishes in
one direction in momentum space.

We start from \eqn{equ:spinwaveH2} and do the Fourier transform of
the bosonic fields,
\be
b_{X,\veck}=\cellno^{-1/2}\sum_{\vecR}{\exp[-\im \veck\cdot(\vecR+\vecr_X)]
b_{X,\vecR}}
\nn\label{equ:bFT}
\ee
where $X={\rm A,B,C}$ labels the three sublattices, $\cellno$ is the number
of unit cells. $\vecR$ are positions of unit cells, $\vecr_X$ are positions of the
three basis sites within a unit cell, and $\veck$ is the wavevector.

The quadratic Hamiltonian is then block-diagonalized
\be
H_2=\sum_{\veck}{\psi\yd_\veck\cdot M\nd(\veck)
\cdot \psi\nd_\veck+{\rm constant}}
\nn
\ee
where $\psi\yd_\veck=(b\yd_{A,\veck},b\yd_{B,\veck},b\yd_{C,\veck},
b\nd_{A,-\veck},b\nd_{B,-\veck},b\nd_{C,-\veck})$, $M(\veck)$ is a $6\times 6$
hermitian matrix, and the summation is over the $\veck$-points in the BZ.
Here $M(\veck)$ can be written as
\be
M(\veck)=\begin{pmatrix}P & Q\\
Q & P
\end{pmatrix}
\nn\label{equ:spinwaveMmatrix} \ee
Here $P$ and $Q$ are both $3\times 3$
matrices as shown below, and we use the notation $c_1 =
\cos(\frac{k_1}{2})$, $c_2 = \cos(\frac{k_2}{2})$, and $c_3 =
\cos(\frac{k_3}{2})$ with $k_i=\veck\cdot \vece_i$,
$k_3=-k_1-k_2$.
\be
\begin{split}
P&=\frac{1}{2\alpha}\begin{pmatrix}
4 & (2\alpha-1)c_3 & (2\alpha-1)c_2\\
(2\alpha-1)c_3& 4\alpha^2 & c_1\\
(2\alpha-1)c_2 & c_1& 4\alpha^2
\end{pmatrix}\\
Q&=\frac{2\alpha+1}{2\alpha}\begin{pmatrix}
0 & c_3 & c_2\\
c_3 & 0 & (2\alpha-1)c_1\\
c_2 & (2\alpha-1)c_1 & 0
\end{pmatrix}
\end{split}
\nn\label{equ:spinwavePQmatrix}
\ee


We need to further diagonalize $M(\veck)$ by an SU(3,3) Bogoliubov
transformation. Namely we need an SU(3,3) matrix $U$ such that
\be
U\yd\tau U\nd=\tau,\quad\tau=\begin{pmatrix}\idmat_{3\times 3} & 0\\
0 & -\idmat_{3\times 3}\end{pmatrix}
\nn
\ee
and
\be
U\yd M(\veck) U\nd=\begin{pmatrix}\omega(\veck) & 0\\
0 & \omega(-\veck) \end{pmatrix}
\nn
\ee
where $\idmat_{3\times 3}$ is the $3\times 3$ identity matrix,
$\omega(\veck)$ is a $3\times 3$ diagonal matrix with three branches of
spin dispersions as the diagonal elements, because of the inversion symmetry
$\omega(\veck)=\omega(-\veck)$.

In isotropic Kagom\'e model $P$ and $Q$ commute and can be
diagonalized simultaneously, which simplifies the calculation. But
for general $\alpha$ matrices $P$ and $Q$ do not commute.

A simpler way to get the dispersion is to solve the eigenvalues of
$\tau\cdot M(\veck)$. It is fairly simple to prove that the six eigenvalues
of $\tau\cdot M(\veck)$ are $\pm\omega_i(\veck)$, $i=1,2,3$ indicating three
branches\cite{Sachdev:}. The characteristic polynomial of $\tau\cdot M(\veck)$
is $x^6-2f_2 x^4+f_4 x^2$. The dispersion is the following
\be
\omega_1=0,\quad \omega_{2,3}=\sqrt{f_2\mp\sqrt{\Delta}}
\nn
\ee
where $\Delta=f_2^2-f_4$ and
\be
\begin{split}
f_2=&2\alpha^2+1-2\alpha^{-1}+2\alpha^{-2}-(2\alpha^2-1)\cos(k_1)\\
&-\alpha^{-1}[\cos(k_2)+\cos(k_3)]\\
\Delta=&2\frac{(\alpha-1)^2}{\alpha^4}\\
&\times\{2+\alpha^2+\alpha^2\cos(k_1)-2\alpha[\cos(k_2)+\cos(k_3)]\}
\end{split}
\nn \ee Although the dispersion has become much more complicated
than the Kagom\'e case, the zero-energy band still exists.

When $\alpha=1$, $f_2^2-f_4=0$, we have $\omega_2=\omega_3=
\sqrt{3-\cos(k_1)-\cos(k_2)-\cos(k_3)}$. For small $|\veck|$ the dispersion
becomes $\omega_2=\omega_3\sim \sqrt{k_1^2+k_1 k_2+k_2^2}\propto |\veck|$.
Thus we have two `linear' spin wave branches.

However, as long as $\alpha\neq 1$, we have $\omega_2\neq \omega_3$ and
$\omega_3(\veck=0)=2|1-\alpha^{-1}|>0$. We still have one Goldstone mode
because $\omega_2(\veck=0)=0$. But the small wavevector dispersion is
drastically changed, $\omega_2 \sim \sqrt{(\alpha^2-1/4)k_1^2} \propto |k_1|$.
Namely the `spin wave velocity' in the $k_2$ direction (vertical direction
in $\veck$-space) vanishes.

\section{Classical spin wave: quadratic theory and chirality interactions}
\label{app:classicalspinwave}

Let us start from \eqn{equ:H2H3H4}, replace $\flucy$ and $\flucz$ by $\rfy$
and $\rfz$, and do the Fourier transforms of $\rfy$ and $\rfz$ (see the previous
appendix for notation) \be
\begin{split}
\rfy_{X,\veck}&=\cellno^{-1/2}\sum_{\vecR}{\exp[-\im \veck\cdot(\vecR+\vecr_X)]
\rfy_{X,\vecR}}\\
\rfz_{X,\veck}&=\cellno^{-1/2}\sum_{\vecR}{\exp[-\im \veck\cdot(\vecR+\vecr_X)]
\rfz_{X,\vecR}}
\end{split}
\nn
\ee
The quadratic Hamiltonian can be block-diagonalized
\be
\begin{split}
\tilH^y_2&=\sum_{\veck}{\chi_{\veck}\yd M_y\nd(\veck) \chi_{\veck}\nd}\\
\tilH^z_2&=\sum_{\veck}{\phi_{\veck}\yd M_z\nd(\veck) \phi_{\veck}\nd}
\end{split}
\nn \ee where $\chi_{\veck}\yd=(\rfy_{{\rm A},-\veck},\rfy_{{\rm
B},-\veck},\rfy_{{\rm C},-\veck})$ and $\phi_{\veck}\yd=(\rfz_{{\rm
A},-\veck},\rfz_{{\rm B},-\veck},\rfz_{{\rm C},-\veck})$, and
$M_{y,z}(\veck)$ are both $3\times 3$ matrices, shown below, where
we use the notation $c_1 = \cos(\frac{k_1}{2})$, $c_2 =
\cos(\frac{k_2}{2})$, and $c_3 = \cos(\frac{k_3}{2})$.
\be
\begin{split}
M_y(\veck)&=\alpha^{-1}\begin{pmatrix}2  & -c_3 & -c_2\\
-c_3 & 2\alpha^2 & (1-2\alpha^2)c_1\\
-c_2 & (1-2\alpha^2)c_1 & 2\alpha^2
\end{pmatrix},\\
M_z(\veck)&=2\begin{pmatrix}\alpha^{-1} & c_3 & c_2\\
c_3 & \alpha & \alpha c_1\\
c_2 & \alpha c_1 & \alpha
\end{pmatrix}
\nn
\end{split}
\ee
It is straightforward to check that
$M_z(\veck)$ has a zero eigenvalue with (not normalized) eigenvector
$(\alpha \sin(k_1 / 2),\sin(k_2 / 2),\sin(k_3 / 2))$ for all $\veck$;
and $M_y(\veck=0)$ has a zero eigenvalue with eigenvector $(1,1,1)$.

For small $|\veck|$ the dispersion of the lowest branch of $M_y(\veck)$ is
approximately $(1/6\alpha)(\alpha^2 k_1^2+k_1 k_2+k_2^2)$.

\begin{figure}
\includegraphics{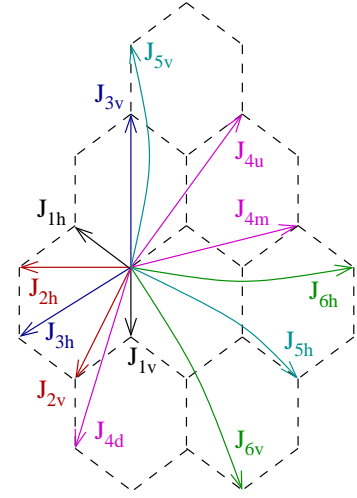}
\caption{(Color online) chirality-chirality couplings calculated here.
Equivalent couplings under space group symmetry are not shown.}
\label{fig:chirchircouplings}
\end{figure}

\begin{table}
\caption{Effective chirality couplings $J_{1\cdots 6}$ (see
\fig{fig:chirchircouplings}) divided by $T^2$, for $J^z=0.1,\
J^y=0.01$. Positive number means antiferromagnetic coupling.}
\begin{tabular}{|r|rrrrr|}
\hline\hline
$\alpha$ & $J_{1v}$ & $J_{1h}$ & $J_{2v}$ & $J_{2h}$ &\\
1 & 0.9702 & 0.9702 & 0.2614 & 0.2614 &\\
1.5 & 1.8231 & -0.3340 & -0.01294 & 0.8895 &\\
\hline
$\alpha$ & $J_{3v}$ & $J_{3h}$ & $J_{4u}$ & $J_{4m}$ & $J_{4d}$ \\
1 & 0.1916 & 0.1916 & 0.002661 & 0.002661 & 0.002661\\
1.5 & 0.4318 & 0.3631 & 0.04897 & -0.2770 & 0.007420\\
\hline
$\alpha$ & $J_{5v}$ & $J_{5h}$ & $J_{6v}$ & $J_{6h}$ &\\
1 & 0.002924 & 0.002924 & 0.002914 & 0.002914 &\\
1.5 & -0.06089 & 0.01172 & -0.03724 & 0.1996 &\\
\hline\hline
\end{tabular}
\label{table:chircouplings}
\end{table}

Now we consider the calculation of the chirality interactions. Each
chirality interaction is calculated by thirty-six terms in
$(\tilH_3)^2$, we show here an example in
\fig{fig:trigGSconstraints}. Chiralities $\chir_1$ and $\chir_5$ are
defined on triangles ABC and DEF in the distorted Kagom\'e lattice,
respectively. $\chir_1$ determines the sign of the angles between
spins on ABC sites, $\theta_{\rm AB}=\chir_1\theta_0,\ \theta_{\rm
BC}=-2\chir_1\theta_0,\ \theta_{\rm CA}=\chir_1\theta_0$.
$\theta_{\rm DE},\ \theta_{\rm EF},\ \theta_{\rm FD}$ are determined
in the similar way by $\chir_5$, and $\theta_{ji}=-\theta_{ij}$.
Plug these into \eqn{equ:H2H3H4}, then the relevant terms in
$(\tilH_3)^2$ are $ 2\chir_1\chir_5(h_{\rm AB}+h_{\rm BC}+h_{\rm
CA})(h_{\rm DE}+h_{\rm EF}+h_{\rm FD}) $ where \be
\begin{split}
h_{\rm AB}&=\sin(\theta_0)(\rfy_{\rm A} \xi_{\rm B}-\rfy_{\rm B} \xi_{\rm A})\\
h_{\rm BC}&=\sin(-2\theta_0)(\rfy_{\rm B} \xi_{\rm C}-\rfy_{\rm C} \xi_{\rm B})\\
h_{\rm CA}&=\sin(\theta_0)(\rfy_{\rm C} \xi_{\rm A}-\rfy_{\rm A} \xi_{\rm C})
\end{split}
\nn
\ee
Here we use $\xi_i=[(\rfy_i)^2+(\rfz_i)^2]$, $\theta_0=\arccos(-1/2\alpha)$
and $h_{\rm DE},\ h_{\rm EF},\ h_{\rm FD}$ are obtained by replacing subscripts
ABC by DEF respectively.

According to \eqn{equ:EffectiveFreeEnergy}
the effective chirality-chirality coupling is
$-T^2\av{(h_{\rm AB}+h_{\rm BC}+h_{\rm CA})(h_{\rm DE}+h_{\rm EF}+h_{\rm FD})}_0$.
Expanding this expression, we have thirty-six terms, each of the form
$\av{\rfy_i \xi_j \rfy_k \xi_m}_0$
which can be further expanded into four terms
$\av{\rfy_i(\rfy_j)^2\rfy_k(\rfy_m)^2}_0
+\av{\rfy_i(\rfy_j)^2\rfy_k(\rfz_m)^2}_0
+\av{\rfy_i(\rfz_j)^2\rfy_k(\rfy_m)^2}_0
+\av{\rfy_i(\rfz_j)^2\rfy_k(\rfz_m)^2}_0$.
Each term in the last expression can be expanded into a sum of
products of three two-point correlators by Wick theorem.
The two-point correlators are computed following the standard routine
in all quadratic theory, {\it e.g.}
\be
\av{\rfy_{{\rm A},0}\rfy_{{\rm B},\vecR}}_0=\int{\dif^2\veck [M_y^{-1}(\veck)]_{\rm AB} e^{\im \veck\cdot (\vecR+\vecr_{\rm B}-\vecr_{\rm A})}}
\nn
\ee
for the A-sublattice site in the unit cell at origin and the B-sublattice site
in the unit cell at position $\vecR$.
We calculated up to the sixth neighbor chirality couplings
(\fig{fig:chirchircouplings}). Some data are presented
in \tbl{table:chircouplings}.

All the above mentioned calculations in \app{app:spinwavedispersion}\ and
\app{app:classicalspinwave}  were done by the software Mathematica.


\begin{thebibliography}{}
\bibitem{note1} An experimental caveat here is that no analogous
signal is seen in $\mu$SR at the same temperatures\cite{Fukaya:},
perhaps due to different local environments of the two probes.
\end{thebibliography}
\end{document}